\documentclass[twocolumn]{emulateapj}
\usepackage{hyperref}
\usepackage{natbib}
\usepackage{comment}

\def\ba{\begin{eqnarray}}
\def\ea{\end{eqnarray}}
\usepackage{graphicx, amsmath, amsthm, amssymb,color}

\shorttitle{Dynamically hot super-Earths from outer giant planet scattering}
\shortauthors{Huang et al.}

\begin{document}

%%%%%%%%%%%%%%%%%%%%%%%%%%%%%%%%%%%%%%%%%%%%%%%%%%%%%%%%%%%%
% TITLE %
%%%%%%%%%%%%%%%%%%%%%%%%%%%%%%%%%%%%%%%%%%%%%%%%%%%%%%%%%%%%
%\title{Dynamical imprints of outer giant planets on Close-in super-Earths: production 
%of eccentric and inclined singles}

\title{Dynamically hot super-Earths from outer giant planet scattering}

%%%%%%%%%%%%%%%%%%%%%%%%%%%%%%%%%%%%%%%%%%%%%%%%%%%%%%%%%%%%
% AUTHOR %
%%%%%%%%%%%%%%%%%%%%%%%%%%%%%%%%%%%%%%%%%%%%%%%%%%%%%%%%%%%%
\author{Chelsea X. Huang\altaffilmark{1,2,3},
Cristobal Petrovich\altaffilmark{3,4}, \& Emily Deibert\altaffilmark{1,2}}

\altaffiltext{1}{Dunlap Institute for Astronomy \& Astrophysics, University of Toronto, 50 St. George Street, Toronto, Ontario, M5S 3H4, Canada}
\altaffiltext{2}{Centre for Planetary Sciences, Department of Physical \& 
Environmental Sciences, University of Toronto at Scarborough, Toronto, 
Ontario M1C 1A4, Canada}
\altaffiltext{3}{MIT Kavli Institute for Astrophysics and Space Research, Cambridge, MA02139, USA, xuhuang@mit.edu}
\altaffiltext{4}{Canadian Institute for Theoretical Astrophysics, University of Toronto, 60 St. George Street, Toronto, Ontario, M5S 1A7, Canada} %\email{cpetrovi@cita.utoronto.ca}

%%%%%%%%%%%%%%%%%%%%%%%%%%%%%%%%%%%%%%%%%%%%%%%%%%%%%%%%%%%%
% ABSTRACT %
%%%%%%%%%%%%%%%%%%%%%%%%%%%%%%%%%%%%%%%%%%%%%%%%%%%%%%%%%%%%
\begin{abstract}

The hundreds of multiple planetary systems discovered by the \textit{Kepler} mission are typically observed to reside in close-in ($\lesssim0.5$ AU), low-eccentricity,  and low-inclination orbits.
We run N-body experiments to study the effect that unstable outer
($\gtrsim1$ AU)  giant planets, whose end orbital configurations resemble those in the Radial Velocity population, have on these close-in multiple super-Earth systems.
Our experiments show that the giant planets greatly reduce the multiplicity of the inner super-Earths and the surviving population can have large eccentricities ($e\gtrsim0.3$) and inclinations ($i\gtrsim20^\circ$) at levels that anti-correlate with multiplicity.
Consequently, this model predicts the existence of a population of dynamically hot single-transiting planets with typical eccentricities and inclinations 
%in the ranges 
of $\sim 0.1-0.5$ and $\sim 10^\circ-40^\circ$.
We show that these results can explain the following observations: (i) the recent eccentricity measurements of \textit{Kepler} super-Earths from transit durations; (ii) the tentative observation that single-transiting systems have a wider distribution of stellar obliquity angles compared to the multiple-transiting systems; (iii) the architecture of some eccentric super-Earths discovered by Radial Velocity surveys such as HD\,125612c.
Future observations from \textit{TESS} will reveal many more dynamically hot
 single transiting planets, for which follow up Radial Velocity studies will be able to test our models and see whether they have outer giant planets.

\end{abstract}

%%%%%%%%%%%%%%%%%%%%%%%%%%%%%%%%%%%%%%%%%%%%%%%%%%%%%%%%%%%%
% KEYWORDS %
%%%%%%%%%%%%%%%%%%%%%%%%%%%%%%%%%%%%%%%%%%%%%%%%%%%%%%%%%%%%

%%%%%%%%%%%%%%%%%%%%%%%%%%%%%%%%%%%%%%%%%%%%%%%%%%%%%%%%%%%%
% INTRODUCTION %
%%%%%%%%%%%%%%%%%%%%%%%%%%%%%%%%%%%%%%%%%%%%%%%%%%%%%%%%%%%%
\section{Introduction} \label{sec:intro}

Originally launched in 2009, NASA's \textit{Kepler} mission \citep{Borucki2010} is responsible for the discovery of thousands of planetary candidates, including over 3000 confirmed planets (e.g., \citealt{Mullally2015,Burke2015,Morton2016}). Through monitoring periodic changes in brightness of light curves from stars (i.e. the ``transit method''), \textit{Kepler} is able to detect planets with radii on the order of 1 $\text{R}_{\oplus}$, although the majority of planets detected are so-called ``super-Earths'' or ``sub-Neptunes'' (with radii $\sim 1.2 - 3 \text{R}_{\oplus}$, \citealt{Burke2015}). Of the thousands of planetary systems discovered by \textit{Kepler} to date, $80\%$ are single-transit systems (i.e. only one planet is observed to transit), while the other $\sim20\%$ consist of 2-7 transiting planets (\citealt{Mullally2015}).

The multi-transit planet systems in the \textit{Kepler} sample populate dynamically cold orbits with low eccentricities and mutual inclinations ($e,i_{\rm m}\ll1$). 
In particular, the eccentricities derived from transit timing variations (TTVs) are typically $\sim0.01$ \citep{Wl13},
while the transit durations of ensembles of multiple transiting planets can be well-fitted with a Rayleigh distribution (Equation \ref{eq:sigma_e}) with mean values of $\bar{e}\sim0.04$ \citep{VA15,xie16} 
and  $\bar{i}_{\rm m}\simeq1-2^\circ$ 
\citep{FM2012,Fabrycky2014}.

In contrast, the properties of the single-transit planet systems seem to be much less certain. 
\citet{Lissauer2011} first noted with preliminary {\em Kepler} data that when modeling the mutual inclination distribution as a Rayleigh function,
they had difficulty reproducing the large observed ratio of single transiting systems to multiple transiting systems. This ``problem" was later on referred to as the ``\textit{Kepler} dichotomy'' in several other studies (\citealt{Johansen2012,HM13,Ballard2016}). 
A wide range of studies into this problem have been undertaken with varying degrees of success \citep[e.g.,][]{Johansen2012,Moriarty2015,Ballard2016,Dawson:2016}, but a consistent picture is still missing. 

This dichotomy might not only be reflected on the derived occurrences between single and multiple planetary systems, but there is tentative evidence, possibly related to the occurrences through the planetary mutual inclinations, that at least a fraction of the single transiting planets in \textit{Kepler} occupy dynamically hotter orbits.
First, \citet{xie16} find that the best fit to the transit durations of single transiting planets is a single Rayleigh distribution with  $\bar{e}\simeq0.3$. Second, by combining measurements of the star's rotation period, radius, and projected rotational velocity, \citet{MW14} found statistically significant evidence that multiple transiting systems to have lower stellar obliquity angles than their single transiting counterparts. This trend can be indicative of larger individual inclinations in the singles assuming that the initial invariable plane, where planets form, nearly coincides with the host star's equator.

\begin{figure}[htbp!]
\centering
\includegraphics[width=\columnwidth]{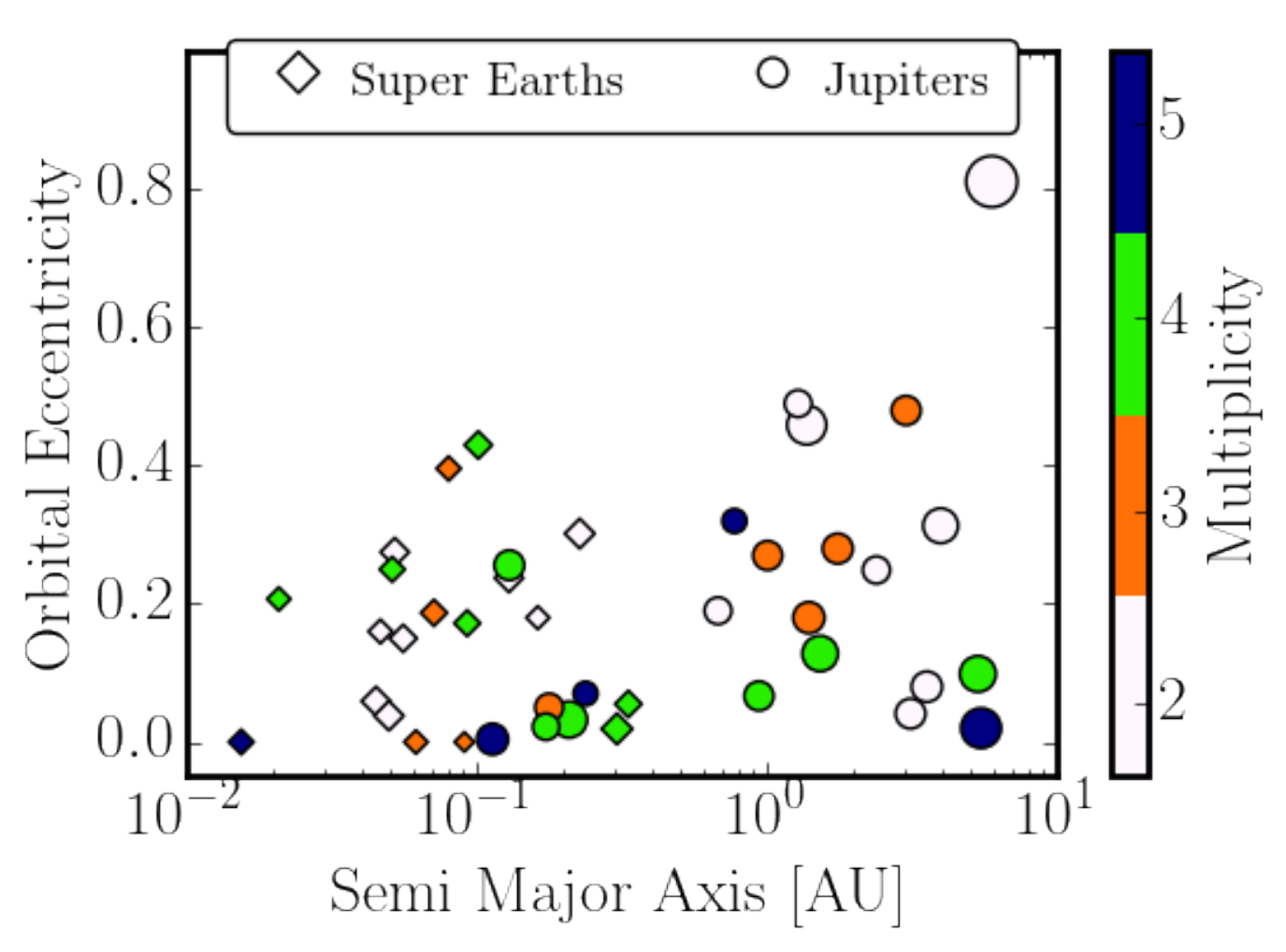}
    \caption{Known planet systems (from exoplanets.org) with long period giant planets and close-in super-Earths ($M{\rm sin}i<0.1\,M_{J}$), color coded by the number of planets in the system. The super Earths (giant planets) are plotted with diamonds (circles), regardless of their period. The size of the markers are proportion to $M{\rm sin}i^{1/3}$. The planet systems we used in this figure including 55 Cnc \citep{Fischer:2008}, BD -08 2823 \citep{Hebrard:2010}, GJ\,832 \citep{Wittenmyer:2014}, GJ\,876 \citep{Correia:2010}, HD\,11964 \citep{Wright:2009}, HD\,125612 \citep{LoCurto:2010}, HD\,181433 \citep{Bouchy:2009}, HD\,190360 \citep{Courcol:2015}, HD\,215497 \citep{LoCurto:2010}, HD\,219828 \citep{Santos:2016}, HD\,47186 \citep{Bouchy:2009}, HIP\,57247 \citep{Fischer:2012}, Kepler-68 \citep{Marcy:2014}, Kepler-89 \citep{Weiss:2013}, and mu Ara \citep{Pepe:2007}.  
\label{fig:observed}}
\end{figure}

A dozen of these dynamically hot close-in super-Earth/Neptune systems have also been discovered in the Radial Velocity surveys, often with giant planets companion further out ($a>1\,AU$). As shown in Figure \ref{fig:observed}, these inner super-Earths (defined as $M\sin i<0.1 M_J$) can easily have reported eccentricity of about $\sim0.1-0.4$. For example, HD\,125612c ($P\sim4$\,day, $M\sin\,i\sim18\,M_{\oplus}$) was determined to have an eccentricity of $0.27\pm0.12$ \citep{LoCurto:2010}.

In this paper, we put forth a connection between these dynamically ``hot" super-Earths and distant giant planets. We propose a scenario by which scattering between distant planets with properties drawn from Radial Velocity surveys can robustly introduce single, eccentric ($e\gtrsim0.3$) and inclined (stellar obliquity $\gtrsim20^\circ$) super-Earth systems, and therefore account for the observed 
differences between single- and multi-transit systems observed in \textit{Kepler} data.

This paper will proceed as follows. In section \S\ref{sec:sims}, we will discuss the details of the code used to run the simulations, as well as the initial conditions chosen to explore the problem. Our results are presented in section \S\ref{sec:results}, including the effects of different populations and initial conditions used in the simulations. 
During the course of this work, many authors explored various interaction between giant planets and close-in super-Earths \citep[e.g.][]{Lai2016,GF16,hansen2016,MDJ16}. We discuss our result together with these works in section \S\ref{sec:discussion}, and our conclusions are presented in section \S\ref{sec:conclusion}.

%\clearpage

\section{Simulations}
\label{sec:sims}
We run N-body simulations of the evolution of the orbits of inner \textit{Kepler}-like planets and outer giant planets ($M_p>0.3\,M_J$) orbiting a solar-type star. 
Planet–star and planet-planet collisions are assumed to result in momentum-conserving mergers with no fragmentation. Collisions are assumed to happen when the distance between two planets (or a planet and the star) becomes less than the sum of their physical radii. The merged body is assumed to conserve total mass and volume. 
A planet is ejected from the system when the distance from the center of mass exceeds 100\,AU and the eccentricity is larger than 1, or when the distance from the center of mass exceeds 1000\,AU.

\subsection{The code}
\label{sec:code}

We use the publicly available integrator IAS15 
\citep{RS15}, which is a high-order scheme
that is part of the REBOUND package \citep{Rein2011}.
 We justify this choice because we 
are mostly interested in the evolution of dynamically-active 
systems where planets experience several close encounters, and 
the IAS15 integrator can handle close encounters with
high precision.

We include the effects from General Relativity in some 
experiments using the package REBOUNDx with the option 
{\it gr-potential} (Tamayo et al., in prep.), which gives 
the right pericenter precession, but gets mean motion wrong by 
$\mathcal{O}(GM/[ac^2])$.

%%%%%%%%%%%%%%%%%%%%%%%%%%%%%%%%%%%%%%%%%%%%%%%%%%%%%%%%%%%%
% INITIAL %
%%%%%%%%%%%%%%%%%%%%%%%%%%%%%%%%%%%%%%%%%%%%%%%%%%%%%%%%%%%%
\subsection{Initial conditions}
\label{sec:init}
%{\bf TBD: divided up}

We first initialize our planetary systems with either super-Earths (the \textit{Kepler} population) or gas giant planets (Radial Velocity population) so we can match the bulk of their orbital architectures after either population has evolved for $>1$ Myr. We assess the match to the observed orbital properties for the \textit{Kepler}-like systems and the Radial Velocity population in \S\ref{sec:kepler} and \S\ref{sec:jupsonly}, respectively.

\subsubsection{Super-Earth population}

\begin{figure}[htbp!]
\includegraphics[width=\columnwidth]{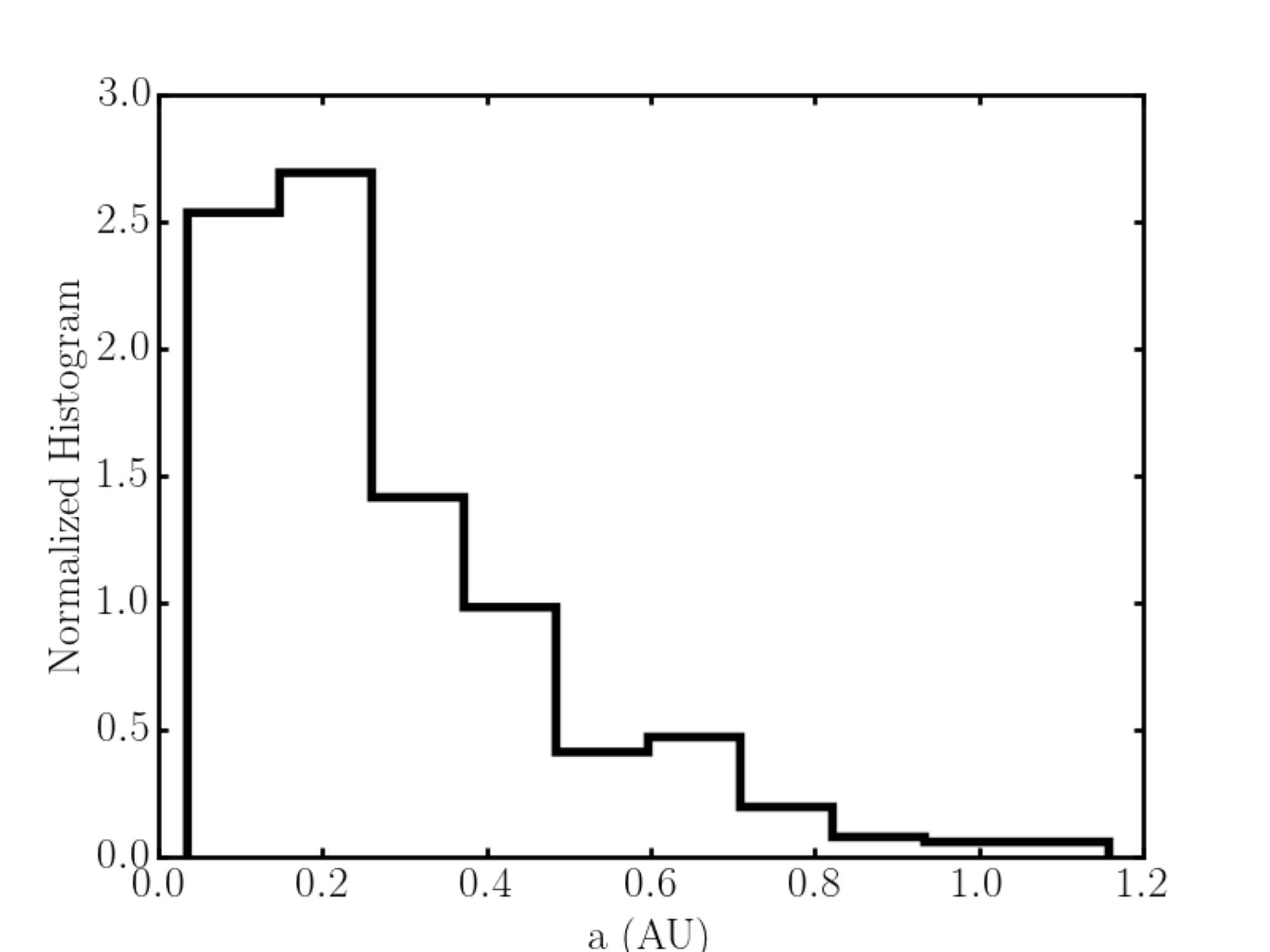}
\includegraphics[width=\columnwidth]{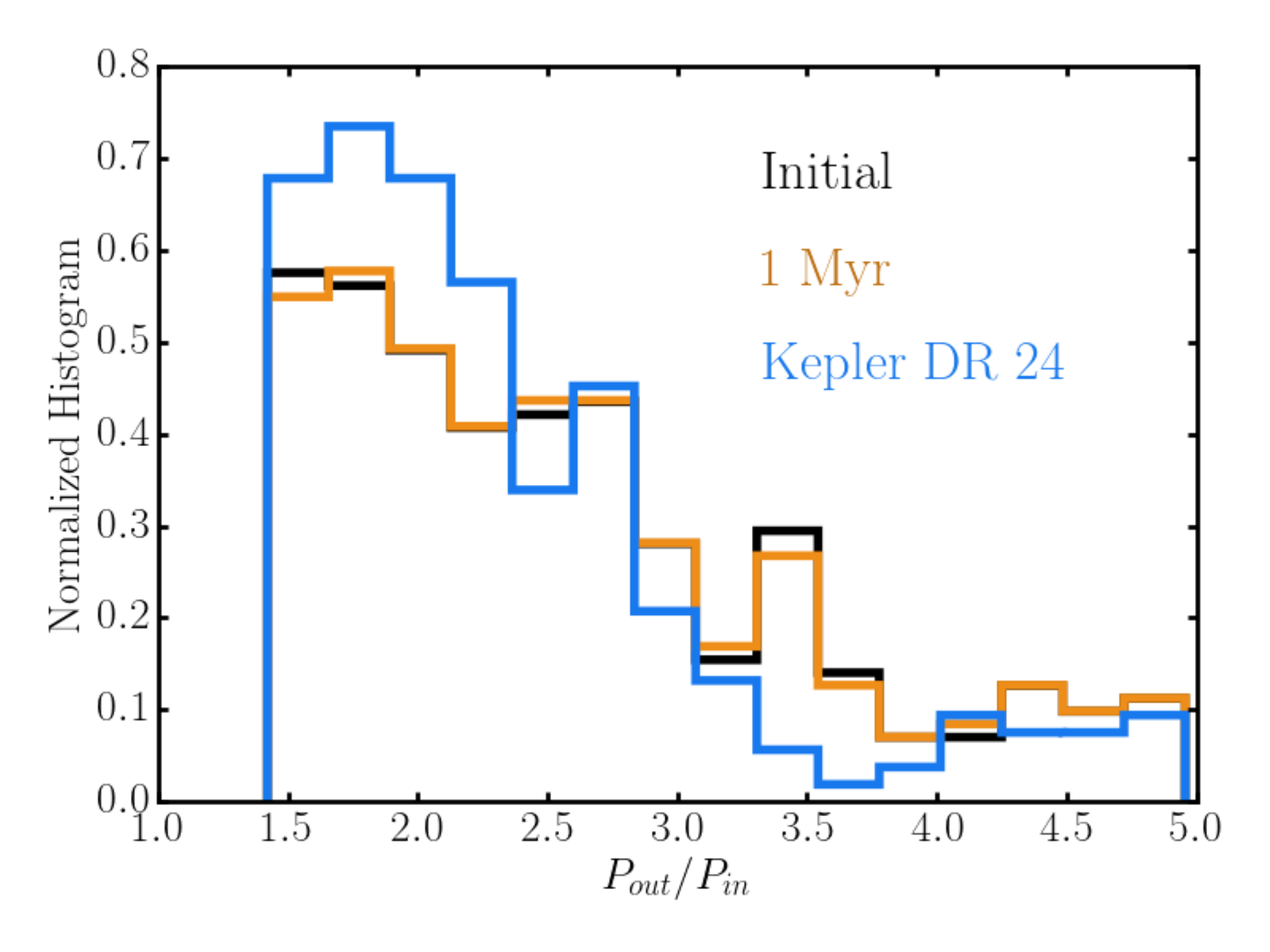}
\caption{The top panel shows the initial semimajor axis distribution of the super-Earths. The bottom panel shows the initial (black) and end (orange) period ratio distribution of neighboring super-Earths. For comparison we plot the period ratio of nearest neighbour for the \textit{Kepler} three planet systems from the release DR24 \citep{Coughlin:2016} in blue. We note that only the \textit{Kepler} systems with period ratio between 1.5 and 5 are shown in this figure. 
\label{fig:init-pratio}}
\end{figure}

We initialize three super-Earths located inwards of $\sim1\,$ AU
and in long-term stable orbits. 
We choose their semi-major axes from the fitting function for the probability distribution of semi-major axes in the {\it  Kepler} sample, after accounting for geometric selection effects \citep{TD11}: 
\ba 
dp(a)=0.656 \frac{(a/a_0)^{3.1}} {1+(a/a_0)^{3.6}}
\frac{da}{a}, 
\label{eq:epsilon}
\ea 
where $a_0=0.085\,$AU and the semi-major axes are restricted to
$ a<1.15\,\mbox{AU}$ \footnote{The choice of 1.15\,AU is due to the limited timespan of the Kepler data, following \citep{TD11} Equation 35.}.
We draw three independent random values from this distribution
and compute the period ratio between neighboring planets $\mathcal{P}$
as
\ba
\mathcal{P}=\left(a_{i+1}/a_{i}\right)^{3/2},
\ea
 where $i$ is in order of increasing semi-major axes.
We further limit the period ratio between any outer planet and its neighboring inner planet to be $1.4<\mathcal{P}<5$ in order to avoid having systems that can become unstable in timescales comparable to the ages of the systems by themselves at $\mathcal{P}<1.4$\footnote{For reference, an evenly-spaced three-planet system with masses of $\sim 10M_\oplus$ generally becomes unstable within $\gtrsim10^{10}$ orbits
when their mutual separations are $\lesssim8$ mutual Hill radii (see Eq \ref{eq:delta_a}), which corresponds
to period ratios of $\lesssim1.4$ for  planets (e.g. \citealt{funk10,Pu2015}).} and also systems that are too widely-spaced  ($\mathcal{P}\gtrsim5$) to be easily destabilized by outer giant planets.
The distribution of period ratios between neighboring planets are shown in Figure \ref{fig:init-pratio} and we observe that it roughly matches the observed one from \textit{Kepler}. We note that the observed distribution has not been corrected for the effect of geometric transit probability, therefore it has an excess at the end with shorter period ratios.

The planets have non-zero eccentricities $e$
and inclinations $i$. These are assumed to be randomly distributed following a Rayleigh law, 
\ba
dp=\frac{x\,dx}{\sigma_x^2} \exp\left(-\onehalf x^2/\sigma_x^2\right),
\label{eq:sigma_e}
\ea
where $x=e$ or $i$ and $\sigma_x$ is an input parameter that
is related to the mean, median and rms eccentricity or inclination by $\langle
x\rangle=\sqrt{\pi/2}\sigma_x=1.253\sigma_x$, $\tilde{x}=\sqrt{2\ln2}\sigma_x=1.18\sigma_x$, and $\langle
x^2\rangle^{1/2}=\sqrt{2}\sigma_x=1.414\sigma_x$. {\bf We assume $\sigma_e$ and $\sigma_i$ both to be 0.01.}

 The masses of the super-Earths were either 5, 10, or 15 $\text{M}_{\oplus}$, but the ordering of the masses was randomized. This choice is arbitrary and the range of masses $5-15~\text{M}_{\oplus}$ is characteristic for planets with $\sim2-4~\text{R}_{\oplus}$ (e.g., \citealt{WM14}).

\subsubsection{The gas giant population}

We choose the semi-major axis to be uniformly distributed 
in a defined range 2-5 AU. 
Labeling the planets by subscripts $i$ in order of increasing 
semi-major axis, we impose a minimum initial spacing 
of the orbits given by
\ba
\Delta a_{i,i+1}&\equiv& a_{i+1}-a_{i}>K R_{H,i,i+1},\mbox{where} \nonumber \\
R_{H,i,i+1}&=&\left(\frac{M_i+M_{i+1}}{3 M_\star}\right)^{1/3}
\frac{a_i+a_{i+1}}{2},
\label{eq:delta_a}
\ea
and $R_{H,i,i+1}$ is the mutual Hill radius of planets
with masses $M_i$ and $M_{i+1}$.

The initial spacing between orbits mainly changes the 
timescale of onset of dynamical instability
(e.g., \citealt{CWB96}). 
Our choice of $K$ in Equation (\ref{eq:delta_a}) is
empirically guided by the fact that we would like to avoid very 
closely-packed systems which might be subject to gravitational focusing, leading to a spurious excess of planet-planet collisions (e.g., \citealt{PTR14}). 

We initialize the system with three gas giant planets
and use $K=3$.
For reference, a crude estimate of the instability time
can be obtained from the numerical experiments by  
\citet{CFMS2008} using a different initial spacing law $\Delta a_{i,i+1}=\tilde{K} R_{H,i,i+1}$ 
(i.e., the spacing is a fixed multiple of the Hill radius, rather than 
exceeding a multiple of the Hill radius). 
They show that for a distribution of planet masses in the
range $0.4-4~M_J$ the median instability 
timescale can be fitted by the following expression:
\ba
\log_{10} (t/\mbox{orbits})=0.021+0.03\exp(1.1~\tilde{K}),
\label{eq:K_tilde}
\ea
where the orbits are those of the innermost planet. 
By assuming that the spacing is a fixed multiple of the Hill radius and that the planet have Jupiter 
masses, our range of semi-major axes of $2-5$ AU allows for $\tilde{K}$ in the range $\sim 3-5$  meaning instability timescales spans in 
$\sim10-10^7$ orbits. In practice, our experiments 
show that most (96$\%$) systems become unstable within 1 Myr (see the results in \S\ref{sec:jupsonly}). 
The masses of the Jupiters are uniformly drawn between 0.3 $M_{\rm J}$ and 3 $M_{\rm J}$. This range of masses is chosen such that our Jupiter-like planets reproduce the bulk of the eccentricities and semi-major axes of the radial velocity population (see Section \S \ref{sec:jupsonly}). The initial eccentricity and inclination distribution of our simulation has a Rayleigh width of $\sigma_e = 0.01$ and $ \sigma_i= 0.01$.

\begin{figure}[htbp!]
\centering
\includegraphics[width=\columnwidth]{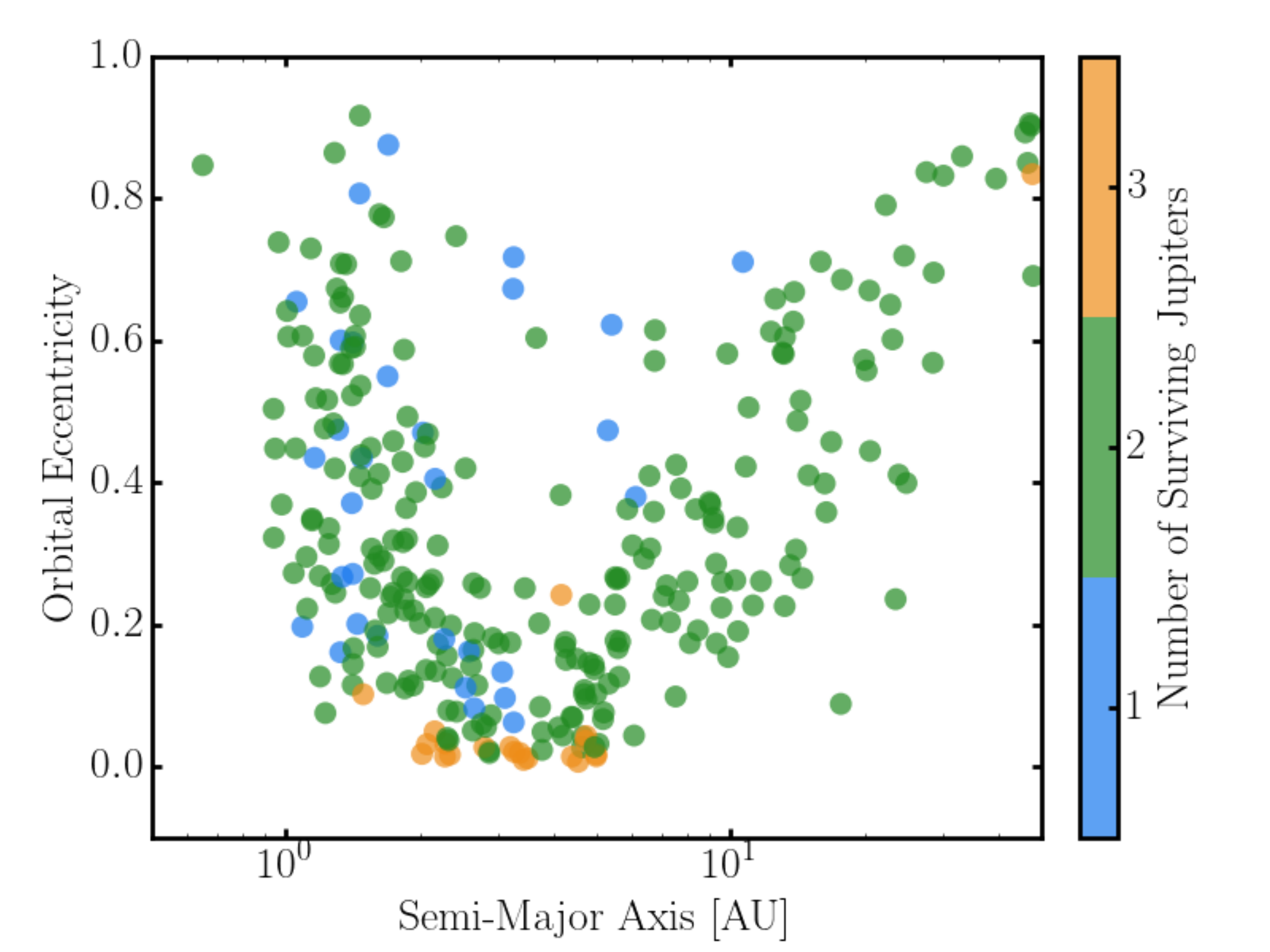}
\caption{Eccentricity versus semi-major axis of the Jupiter-like planet at the end of 1\,Myr evolution. The planets are colour-coded by the multiplicity of their system.}
\label{fig:mass-final}
\end{figure}

\begin{figure}[htbp!]
\centering
\includegraphics[width=\columnwidth]{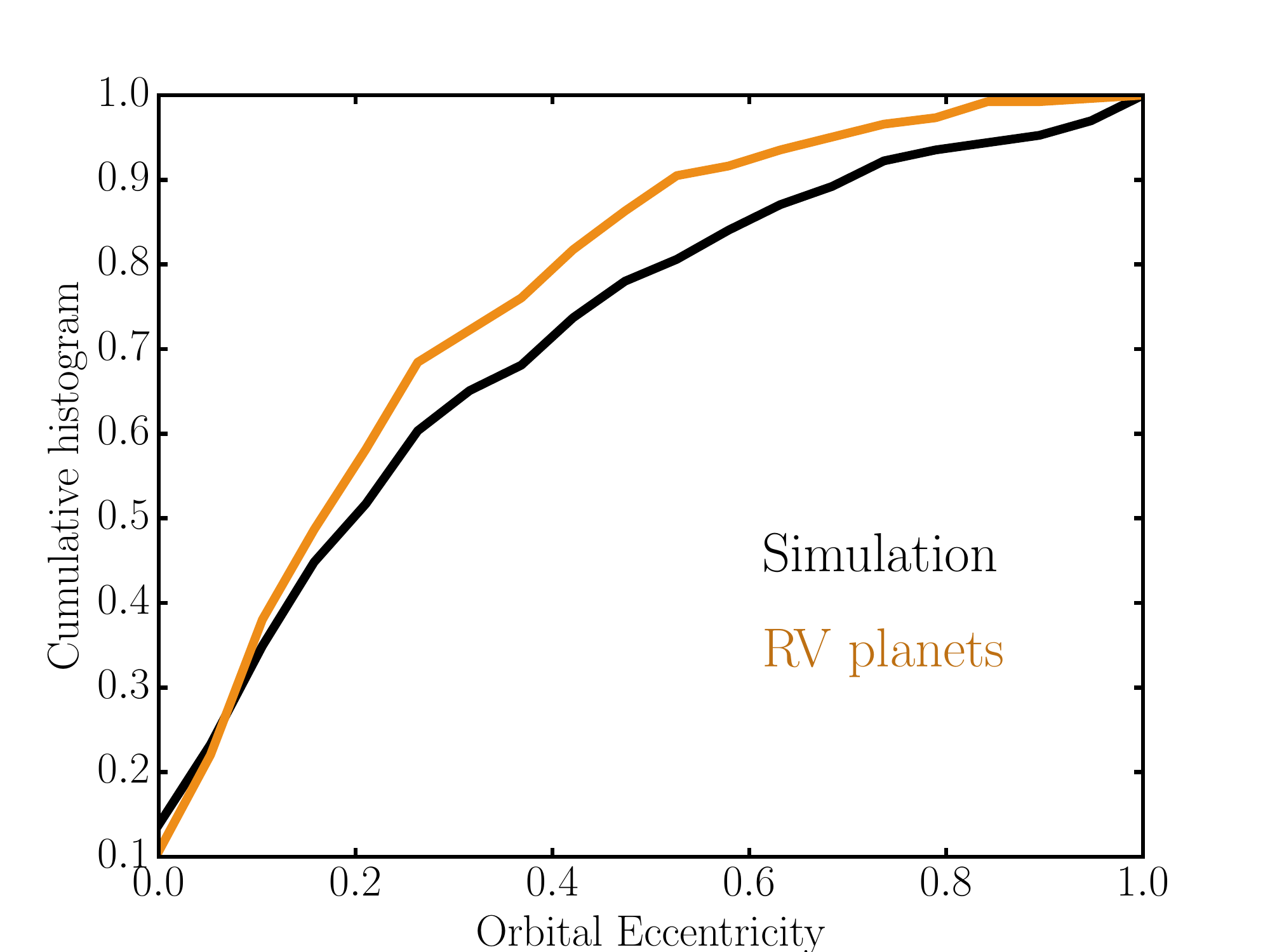}
\includegraphics[width=\columnwidth]{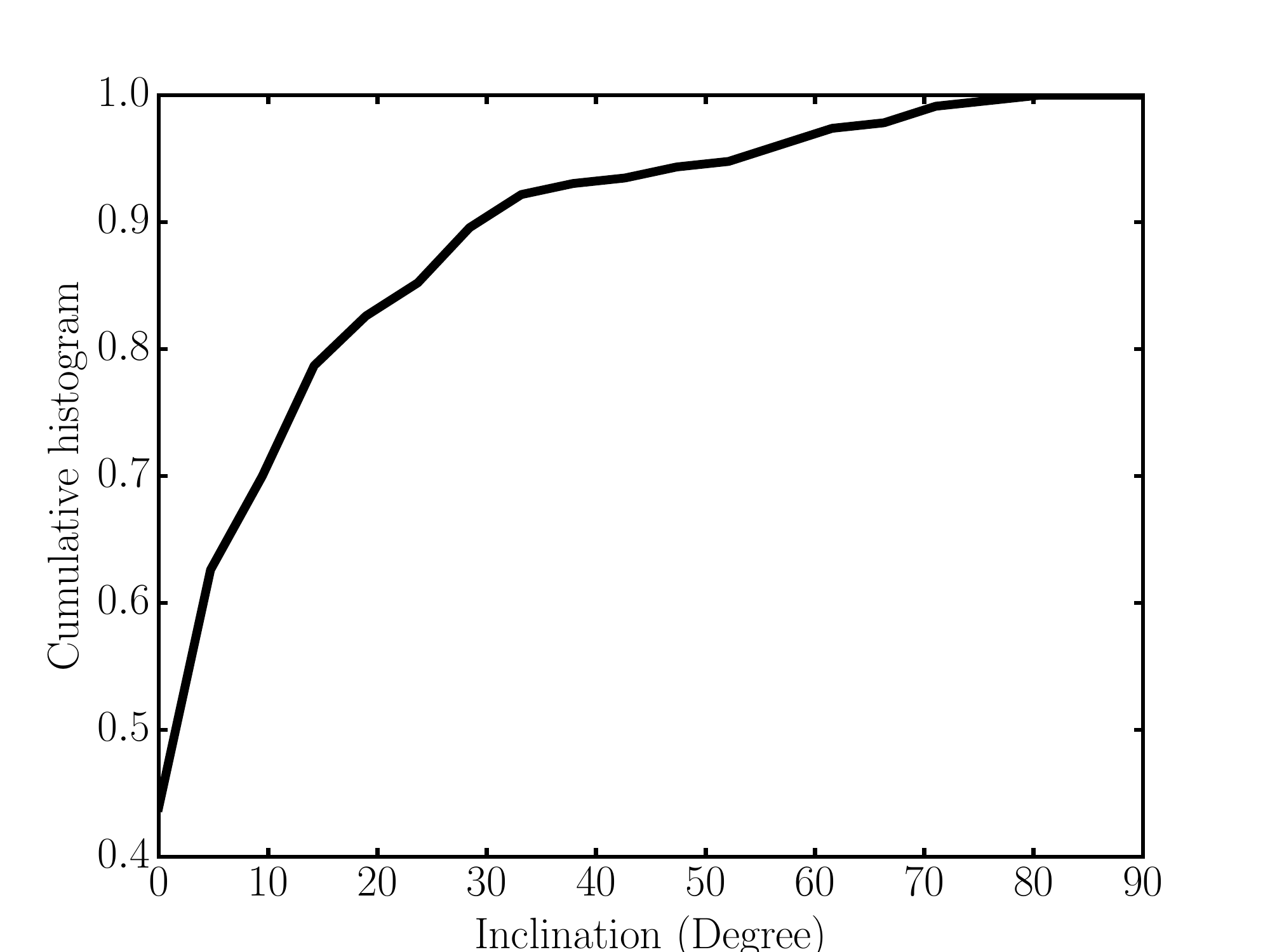}
\caption{The final cumulative eccentricities (upper panel) and inclinations (lower panel) distribution of the Jupiter like planets within 5\,AU after 1\,Myr evolution (black line). Orange line indicate the cumulative eccentricity distribution of planets detected by radial velocity method with $a<5\,AU$.
The initial eccentricity and inclination distribution of our simulation has a Rayleigh width of $\sigma_e=0.01$ and $\sigma_i=0.01$\,rad.
\label{fig:hist-jup-only}}
\end{figure}

\section{Results}
\label{sec:results}
\subsection{The inner Kepler-like population only}
\label{sec:kepler}

We evolve 158 systems for 1 Myr. 
As expected from the initial minimum spacing between the planets (see Footnote 4), we find that all systems remain dynamically stable with only small changes in their orbital elements.
Thus, the eccentricity and inclination distributions at the end of the integration follow a Rayleigh distribution with $\sigma_e,\sigma_i\sim0.01$, roughly reproducing the orbital architecture in the multi-planet systems from {\it Kepler} \citep{Fabrycky2014}.

\subsection{The outer radial velocity population only}
\label{sec:jupsonly}

\begin{table}[htbp!]
\centering
\begin{tabular}{c  c}
\hline
\hline
$\text{N}_{\text{J}}$ & Percentage of Systems (\%) \\
\hline
1 & 19 \\
2 & 76 \\
3 & 4  \\
\hline
\end{tabular}
\caption{Multiplicity distribution of the Jupiter like planets.\label{tab:randommass}}
\end{table}

We evolve 160 realizations of systems for 1\,Myr. 
The fraction of systems with a given final number of planets is shown in Table \ref{tab:randommass}. We observe that most systems ($\sim75\%$) end up with two planets (typically one planet in $\sim1-3$ AU and outer one outside $\sim5$ AU, see Figure \ref{fig:mass-final}). 
The prevalence of two-planet systems is also observed in other similar scattering experiments of three giant giant planets
(e.g., \citealt{CFMS2008,Johansen2012,PTR14}). 
This orbital architecture, namely having one cold Jupiter ($a\sim1-5$ AU) and an exterior planetary companion, is consistent with the results by  \citet{bryan16} from Radial Velocity follow-up measurements and adaptive optics surveys that estimated that about half of the cold giants in their survey have outer planetary companions. 
Also, only $\sim5\%$ of the systems stay relatively inactive, which are expected to become unstable if were evolved for long enough time.

We show the eccentricity distribution of giant planets with $a<5$\,AU in Figure \ref{fig:hist-jup-only}, which can be directly compared to the Radial Velocity sample. 
The mean and median eccentricities of these systems are 0.31
and 0.24, which roughly reproduce the observed values of $\simeq 0.26$ and $0.2$ of planets at $>1$\,AU. 
This is not a coincidence as we have tuned the range of the mass distribution and used $0.3-3M_J$ to explain the observe eccentricities\footnote{The resulting eccentricities after scattering of two planets decreases as their mass ratio $\max\{m_1/m_2,m_2/m_1\}$ increases \citep{Ford2008}.}.

Most of the systems have relatively low inclinations, with a mean and median inclination of $\sim 12.7^\circ$ and    $\sim6^\circ$, respectively (Figure \ref{fig:hist-jup-only}).

\begin{figure*}[htbp!]
\includegraphics[width=1.1\columnwidth]{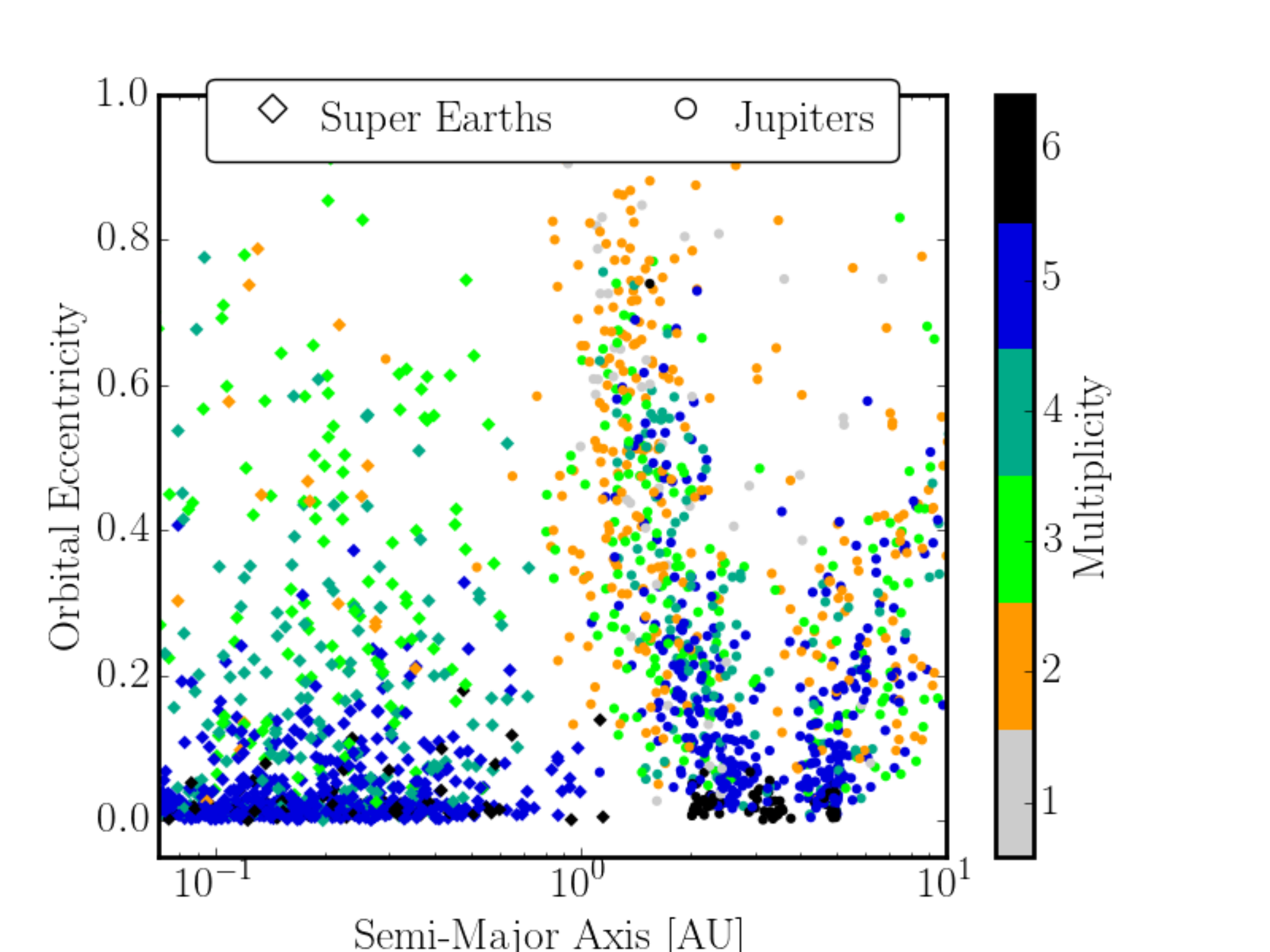}
\includegraphics[width=1.1\columnwidth]{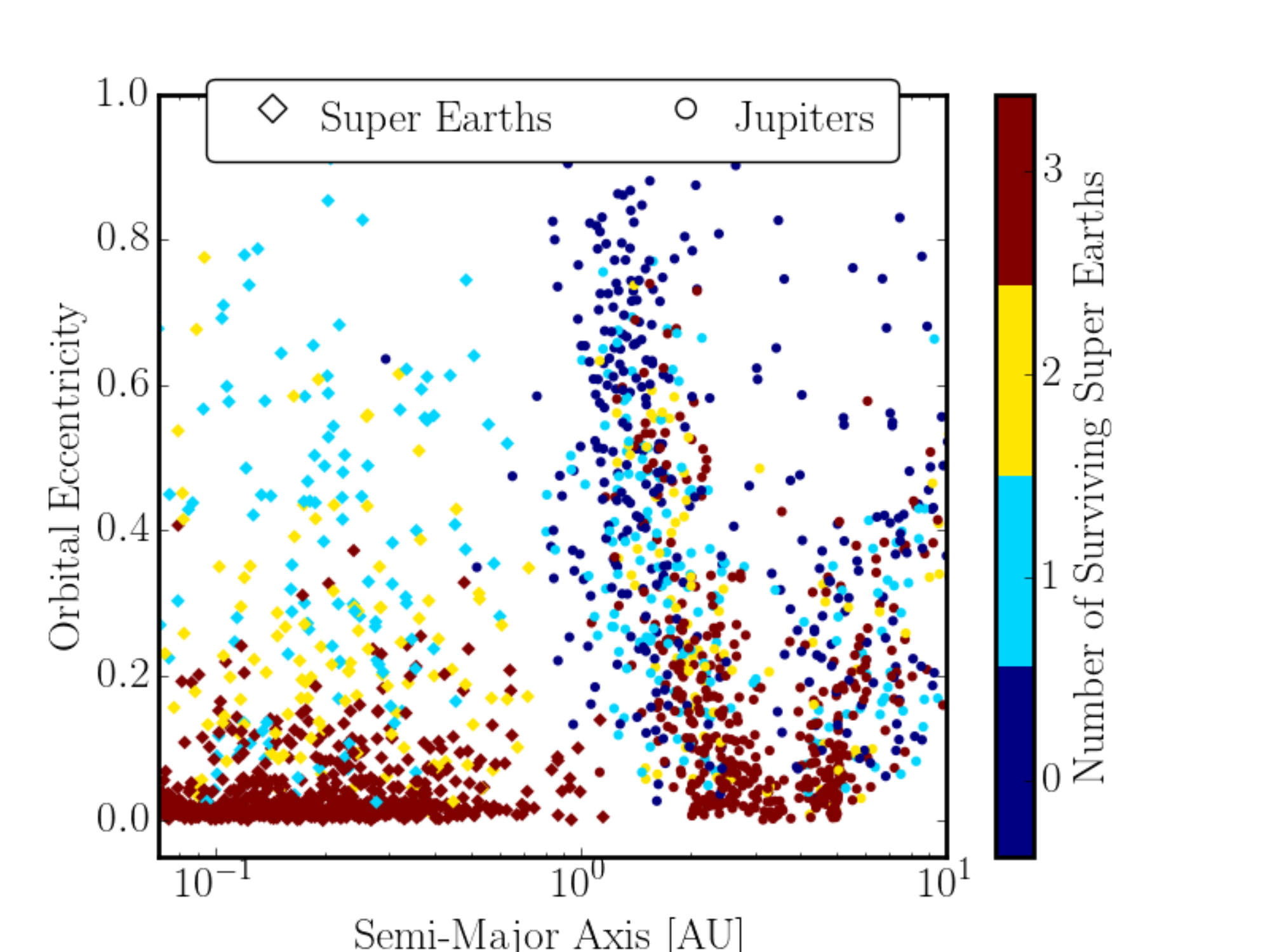} \\
\includegraphics[width=1.1\columnwidth]{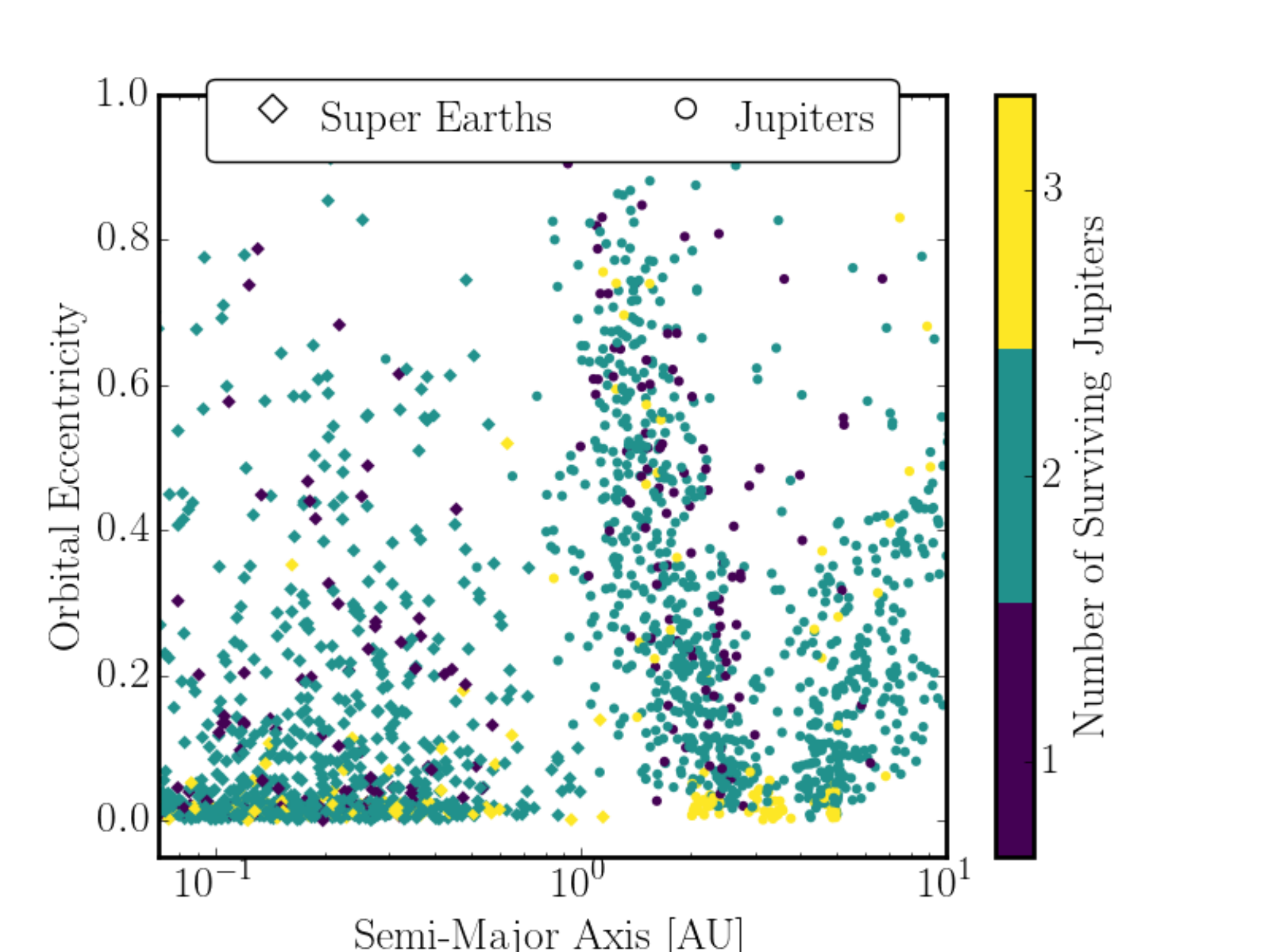}
\includegraphics[width=1.1\columnwidth]{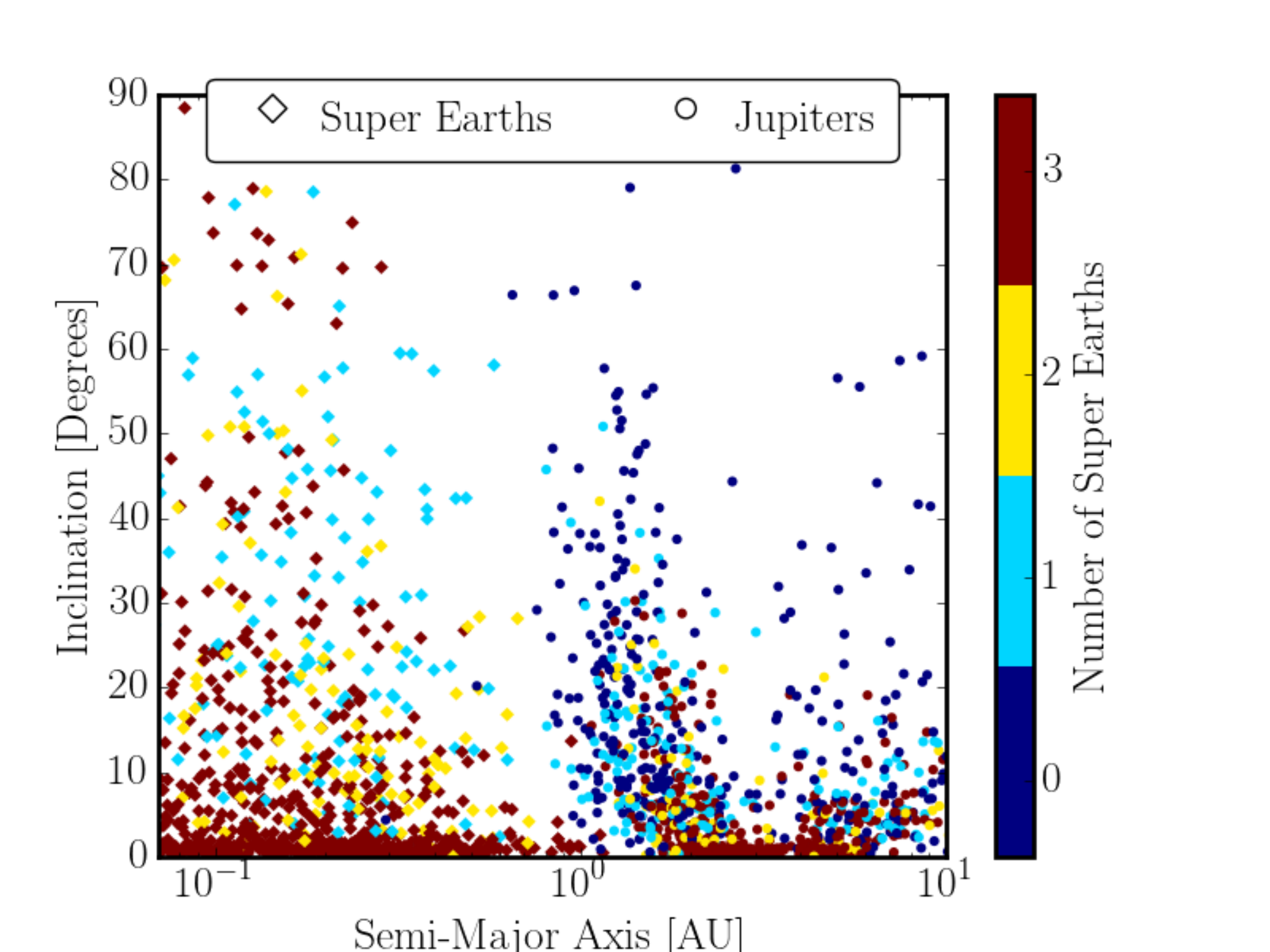}
\caption{
Orbital eccentricity vs semi-major axis for the fiducial run, color-coded by system multiplicity (top left panel), number of super-Earths (top right panel) and number of giant planets (bottom left panel) in the system. 
Bottom right panel shows orbital inclination vs semi-major axis for the fiducial run, color-coded by number of super-Earths in the system.
Jupiter like planets are represented by circles, while super-Earths are represented by diamonds.}
\label{fig:results}
\end{figure*}

%%%%%%%%%%%%%%%%%%%%%%%%%%%%%%%%%%%%%%%%%%%%%%%%%%%%%%%%%%%%
% RESULTS %
%%%%%%%%%%%%%%%%%%%%%%%%%%%%%%%%%%%%%%%%%%%%%%%%%%%%%%%%%%%%
\subsection{The super-Earth and gas giant populations together} \label{sec:standardrun}

Having characterized both the \textit{Kepler} and Radial Velocity populations independently, we now turn to our main experiment, namely putting both populations together.

We first describe the results for our fiducial simulation. In this simulation we set up the systems with initially 3 super-Earths and 3 Jupiters as in the previous sections \S\ref{sec:kepler} and \ref{sec:jupsonly}, while the 
eccentricities and inclinations of all the planets were drawn from Rayleigh distribution in Equation (\ref{eq:sigma_e}) with $\sigma_e=0.01$, and $\sigma_i=0.01$. 

We evolve 653 systems with our fiducial parameters for up 1\,Myr. The outcomes are listed in Table \ref{tab:results}.

In Figure \ref{fig:results}  we show the eccentricity versus semi-major axis of all the remaining planets at the end of 1\,Myr, color-coded by the system's multiplicity, number of remaining super-Earths, number of remaining Jupiters, respectively. The lower-right panel shows the inclination versus semi-major axis of all the remaining planets at the end of 1\,Myr, color-coded by number of remaining super-Earths. We classify the different outcomes as follows: 

\begin{itemize}
\item {\bf No super-Earths:} in slightly less than half of the system ($\sim 40 \%$) the scattering of giant planets fully disrupted the inner super-Earths, leaving mostly two eccentric Jupiter like those in typical Radial Velocity systems. 

\item {\bf Super-Earths with Giant planets:} the remaining half ($\sim55\%$) of the systems lost at least one giant planet due to either ejection or scattering, stir up the close-in super-Earths but do not destroy all of them by either collisions (between planets or between planets and the host star) or ejections.
\begin{itemize}

\item The systems with one, two and three super-Earths remaining together with two giant planets are about 16, 7, and 27 percent of the total outcomes. The remaining single super-Earths almost always have high eccentricities and inclinations. They typically have a flat eccentricity distribution from 0.1 to as high as 0.8, and inclination as high as 60 degrees. The systems with two super-Earths remaining have eccentricity between 0.1-0.4. A fraction of the three super-Earths systems have modest eccentricity between 0.1-0.2, and inclination between 10-20 degree. If a collision happen between the giant planets instead of an ejection, the super-Earths remain dynamically cold. 

\item About $8\%$ of the systems have one eccentric Jupiter left with dynamically hot super-Earths systems. Since the eccentricity distribution of the single Jupiter is similar to those of the inner planet of the two Jupiter systems, the properties of super-Earths are almost indistinguishable from those have two Jupiters.

\end{itemize}

\item {\bf Inactive:} only $4\%$ have all six planets at the end of run. These systems are expected to become unstable in longer time scale. We discuss the effect of integration times in \S\ref{sec:time}.

\end{itemize}

\subsubsection{Eccentricities and inclinations of super-Earths}
\label{sec:e_i_SE}

We take a more detailed look at the orbital configurations of the surviving super-Earths that have lost at least one planet (active systems). 

Hereafter, we shall assume both that the host star spin axis remains fixed throughout the evolution of the systems and that it coincides with the normal to the initial invariable plane. Thus, all the planets have small initial stellar obliquity angles (typically $\lesssim1^\circ$) and the individual inclinations and obliquity angles nearly coincide.

\begin{figure*}[htbp!]
\centering
\includegraphics[width=\linewidth]{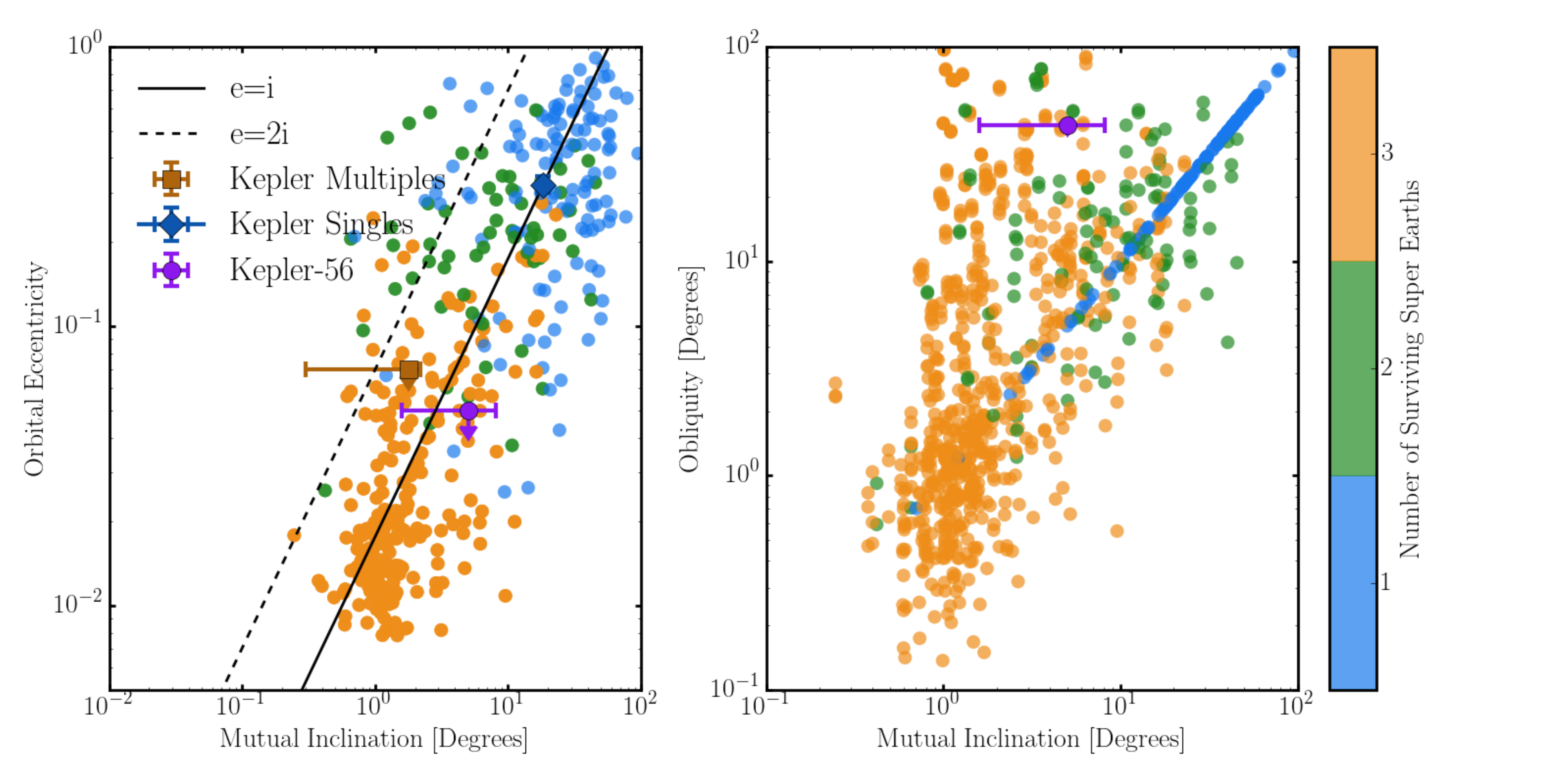}
\caption{Left panel shows mean orbital mutual inclination versus mean eccentricity for the systems with super-Earths survive, color coded by the number of super-Earths in the system. 
We show the linear trend $e=i$ and $e=2i$ in solid and dashed lines to compare with. We also show the location of Kepler 
multiple systems (orange square) and single systems (blue diamond) on this plane as derived by \citet{xie16}. 
The inclination of the Kepler single systems are assumed to be the same as their derived eccentricity.
Kepler-56 system is noted as the purple dot with the inclination taken from the measurement from the gravitational mode of the host star \citep{Huber:2013}, 
and eccentricity taken from the dynamic upper limit estimate from \citet{Huber:2013} assuming a low mutual inclination between the close-in planets. Note that for the three super-Earth systems we show the mean mutual inclination between the three possible pairs of planets. 
Right panel shows the mean mutual inclination of the system versus the stellar obliquity of the individual planets
(angle between the host star spin axis 
and the normal to the planetary obit, which is assumed to be zero initially).  The obliquity of all planets at the start of simulation is assumed to be close to 0. The color coding is the same as the left panel.  
\label{fig:inc_earthonly}
}
\end{figure*}

\begin{figure*}[htbp!]
\centering
\includegraphics[trim=10 0 10 0,width=0.66\columnwidth]{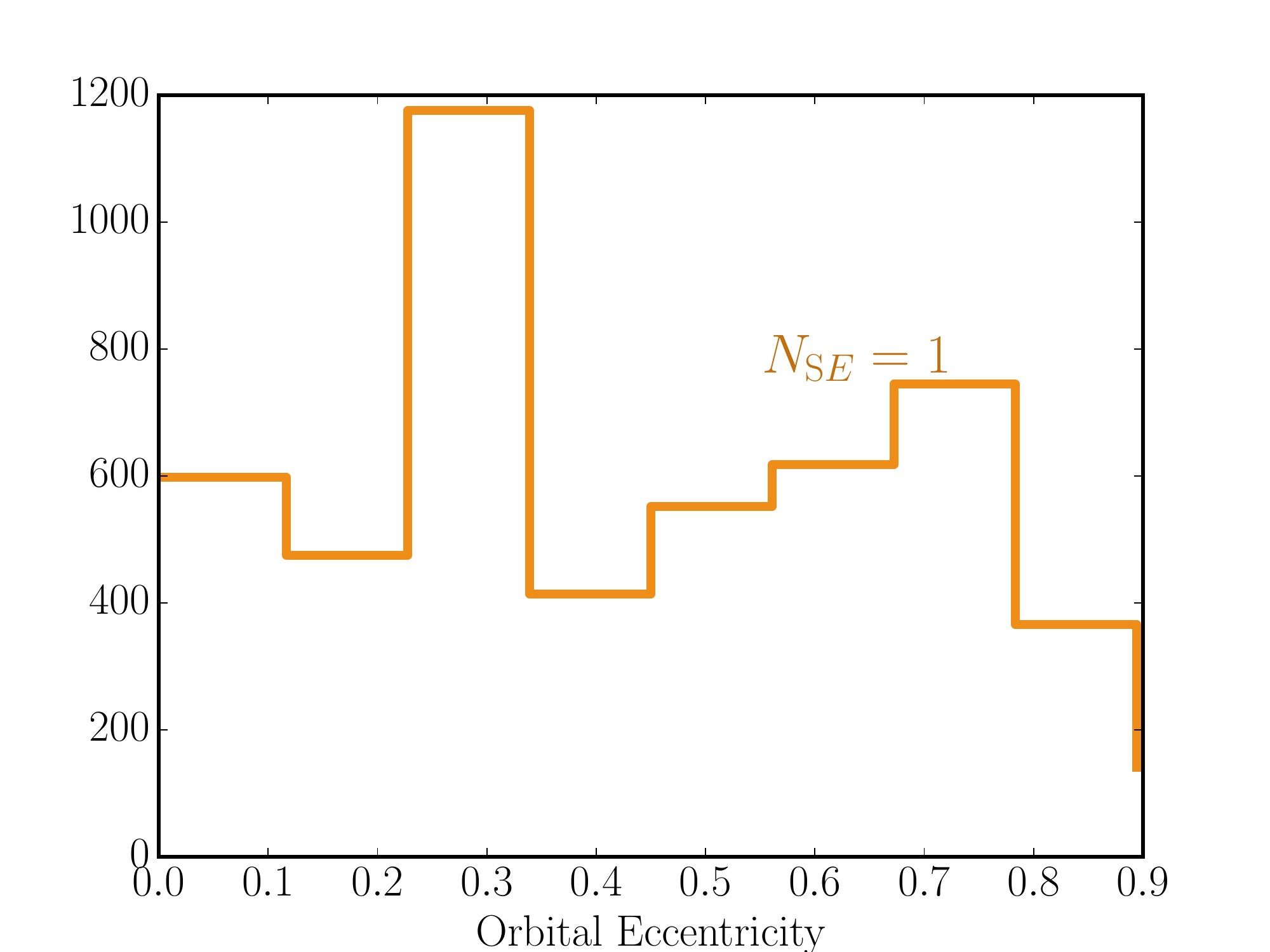}
\includegraphics[trim=10 0 10 0,width=0.66\columnwidth]{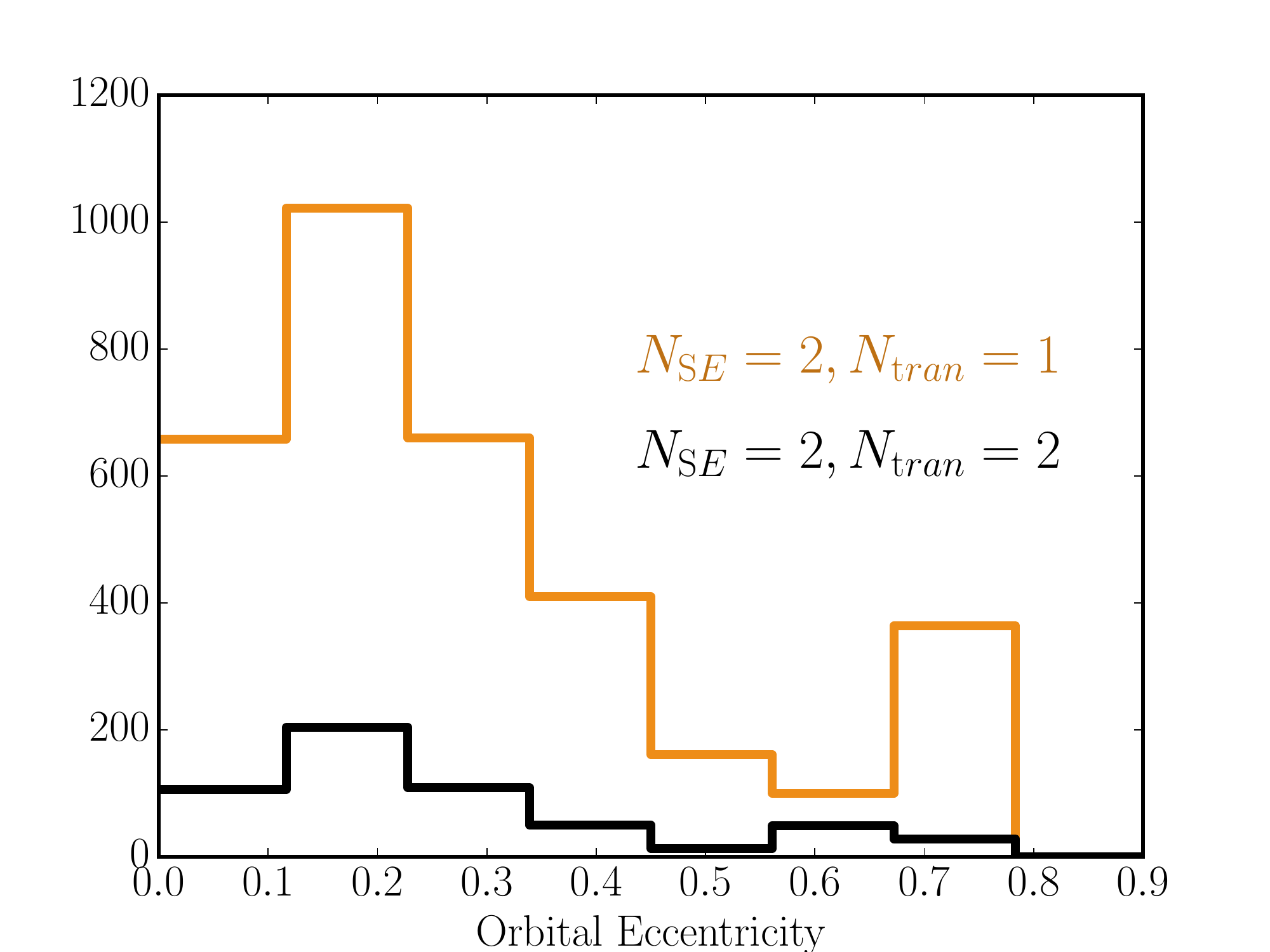}
\includegraphics[trim=10 0 10 0,width=0.66\columnwidth]{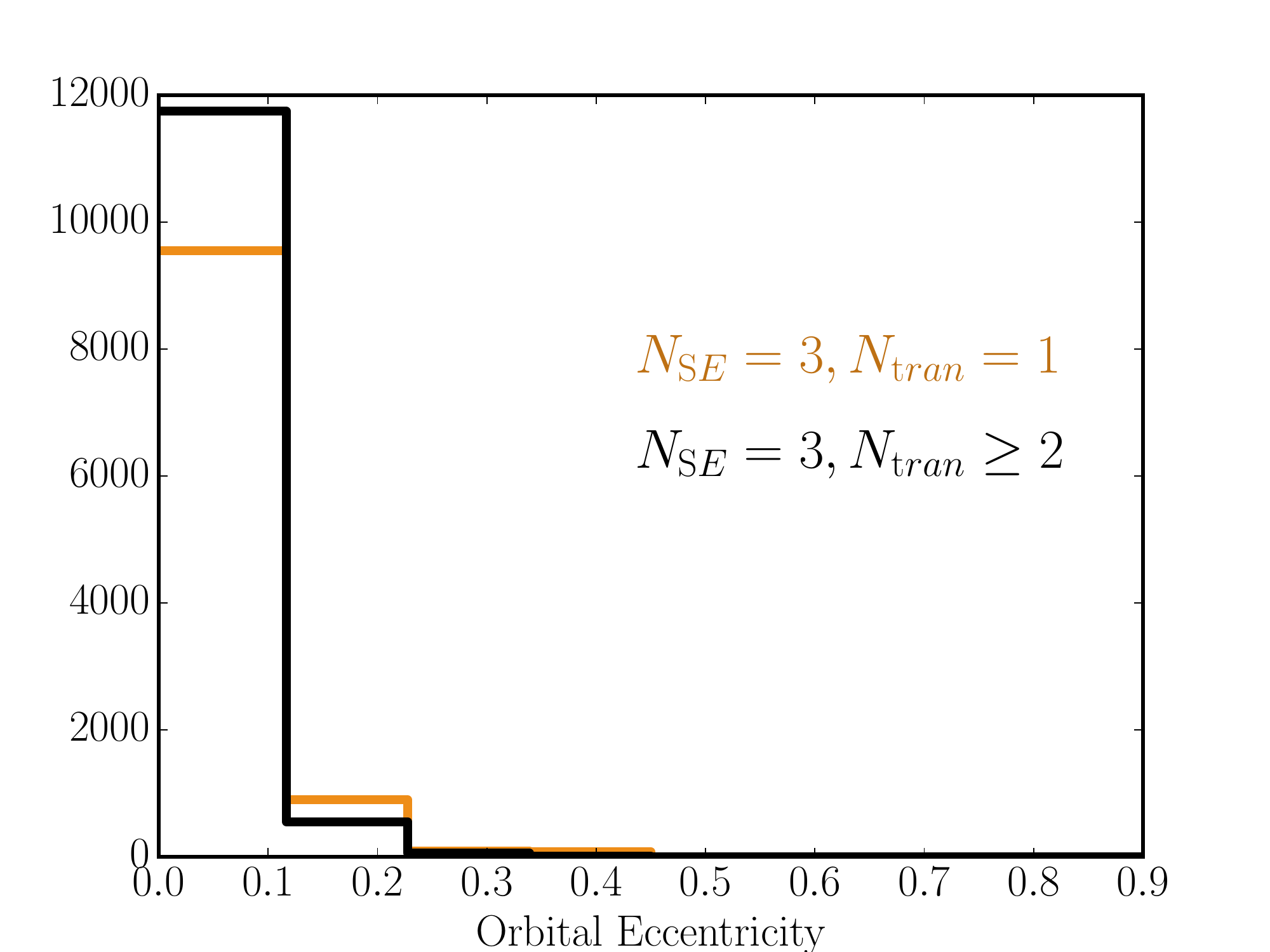}
\includegraphics[trim=10 0 10 0,width=0.66\columnwidth]{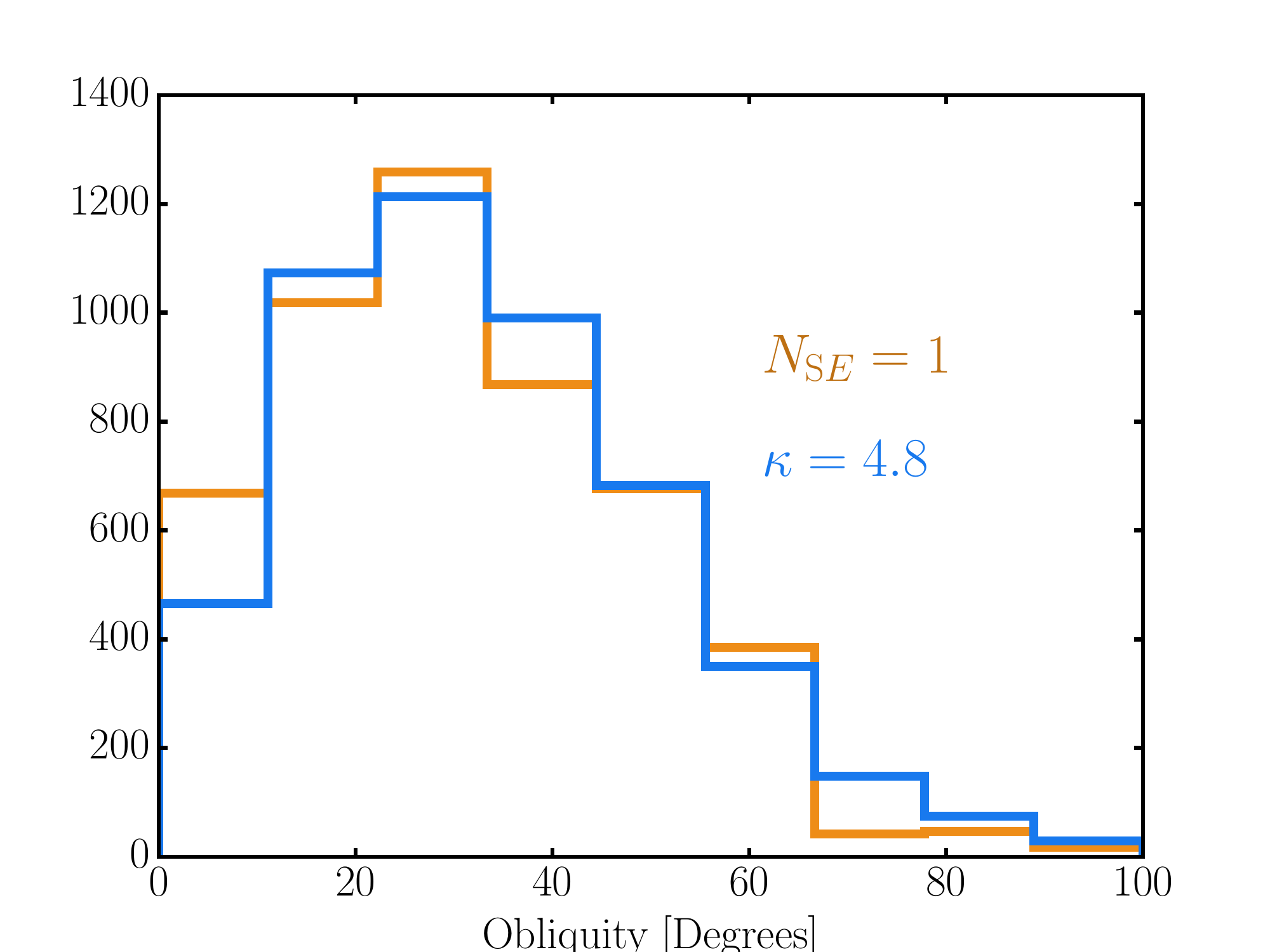}
\includegraphics[trim=10 0 10 0,width=0.66\columnwidth]{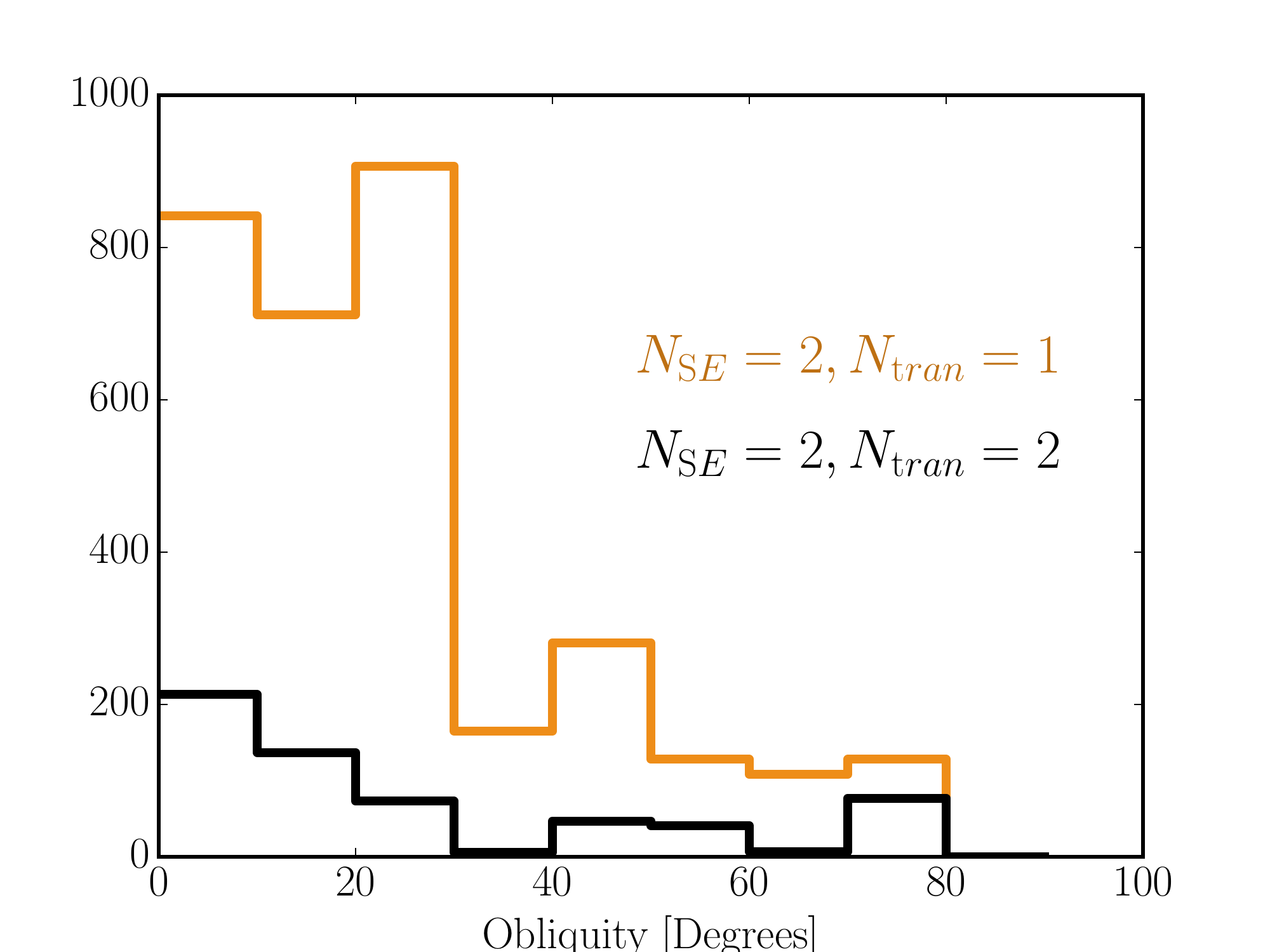}
\includegraphics[trim=10 0 10 0,width=0.66\columnwidth]{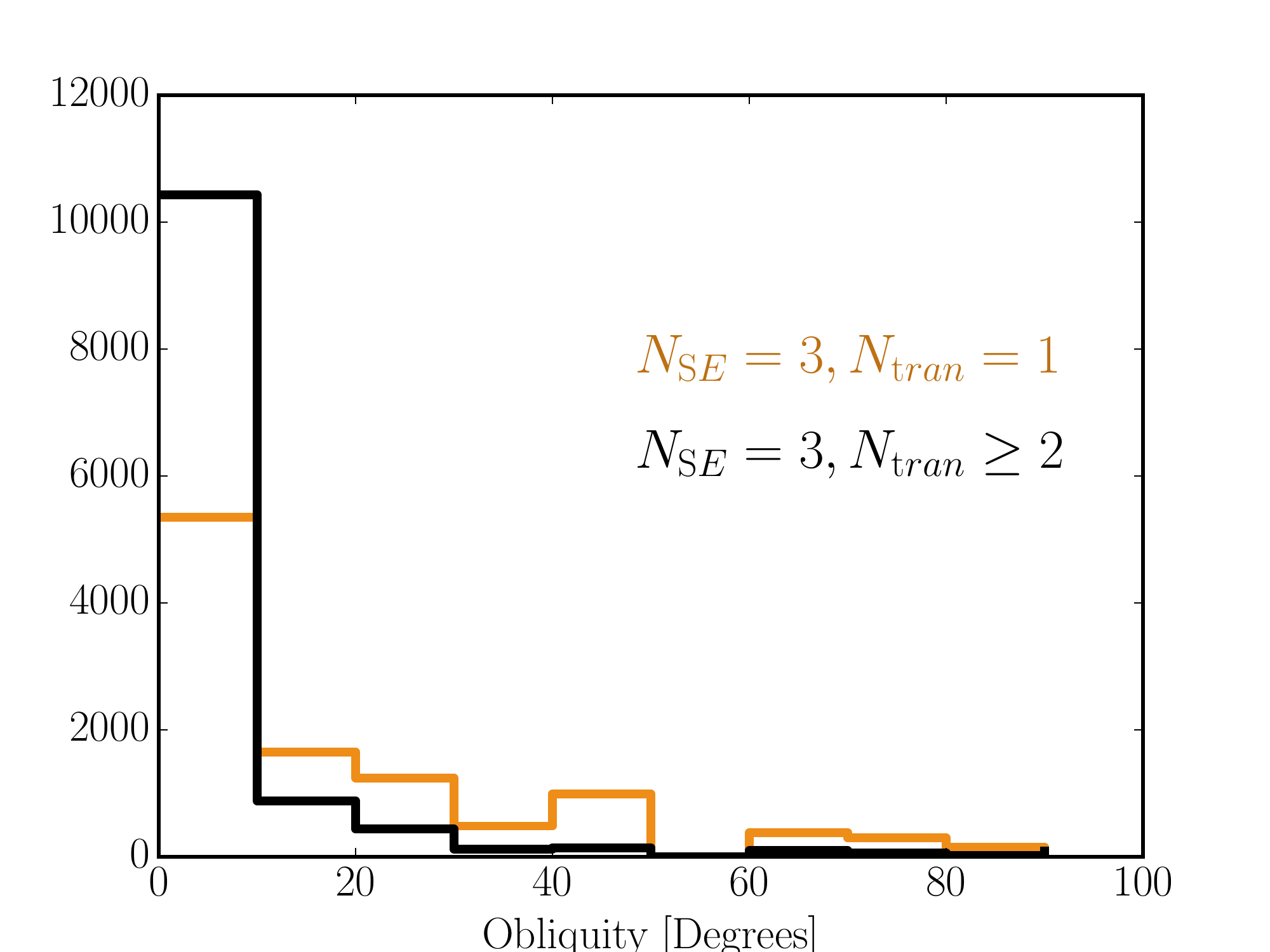}
    \caption{The distribution of eccentricities (upper panels) and  stellar obliquity angles (lower panels) for the super-Earths that survive in our fiducial simulation. From left to right each panel shows the obliquity for systems with 1 to 3 surviving super-Earths. For the systems with $N_{\rm SE}=2$ and $N_{\rm SE} = 3$, we use \texttt{CORBITS} to simulate random observations, and divide them into two samples: systems with one transiting planet (orange lines) and systems with more than one transiting planets (black lines). The distribution representing the best fit Fisher concentration parameter $\kappa$ for the single transit planets fitted by \citet{MW14} is shown in blue lines in the first figure of the lower panel (see section \S\ref{sec:e_i_SE}). 
\label{fig:obl_nearth}}
\end{figure*}

In Figure \ref{fig:inc_earthonly} we show the eccentricities (left panel) and stellar obliquity angles (right panel) as a function of the mutual inclinations of the surviving super-Earths. For the single systems, we use the inclinations of the planets as their mutual inclinations. 
In general, systems with less super-Earths have higher eccentricities and inclinations (both the mutual inclinations and the stellar obliquity angles).  
In Figure \ref{fig:obl_nearth} we show the distribution of eccentricities (upper panels) and  stellar obliquity angles (lower panels) for different numbers of surviving  super-Earths,  $N_{\rm SE}=\{1,2,3\}$. The trend observed in Figure \ref{fig:inc_earthonly} that the systems with higher multiplicity (larger $N_{\rm SE}$) have lower eccentricities and obliquity angles is clearly confirmed. 

We show the comparison of our systems to the \textit{Kepler} single and multiple systems. As derived by \citet{xie16}, the \textit{Kepler} single planet systems can be modeled by an eccentricity distribution with mean of $\bar{e}=0.32\pm0.023$.  We obtain that the mean (median) of the eccentricity and inclination distribution of our single super-Earth population is $\bar{e}=0.39\pm0.02$ ($\tilde{e}=0.35\pm0.04$) and $\bar{i}=30\pm2^{\circ}$ ($\tilde{i}=26\pm2^{\circ}$). 
The mean, median and their 1-$\sigma$ uncertainties is determined by bootstrapping the single planet population for 1000 realizations. This is slightly higher than the estimated value by \citet{xie16}, which it might be expected, since the observed single transit planet systems are likely to be a mixture between the single super-Earths and the multiple super-Earths systems with relatively high mutual inclinations, which have lower eccentricities. 

We use \texttt{CORBITS} \citep{Brakensiek:2016} to simulate random observations of all the systems. \texttt{CORBITS} computes the probability that any particular group of planets can be observed to transit in a multiple planet systems. We assemble 1000 random transit observations for each system and derive the eccentricity and obliquity distribution of single (multiple) transiting systems \footnote{Only the eccentricity and obliquity of the transiting planets are counted in the distribution.}. 

We observe that the systems in the single transiting planets sample have a wider distribution of eccentricity and obliquity. Overall, the sample of single transiting systems have a mean and median eccentricity (obliquity angles) of $0.19\pm0.03$ ($24\pm2^{\circ}$) and $0.09\pm0.02$ ($18\pm2^{\circ}$), respectively.
We caution that the multiple systems with lower mutual inclinations contribute to about 40\% of the distribution. In particular, the single transiting planet systems with three planets survived has mean eccentricity (obliquity) $0.04\pm0.02$ ($19\pm6\,^{\circ}$). The contamination fraction is likely to reduce, since some of the system will become unstable and be disrupted in longer time scales (see Section \ref{sec:time}). Therefore the final distribution will be shifted towards higher eccentricity and obliquity. 

For comparison, the systems with multiple transiting planets have a mean and median eccentricity (obliquity angles) of $0.04\pm0.02$ and $0.02\pm0.01$ ($8\pm5 ^{\circ}$ and $2\pm1 ^{\circ}$). The bulk of these low mutual inclination systems have similar properties as the \textit{Kepler} multiple systems, which are much more circular and coplanar, with an upper limit to the mean eccentricity of $\sim0.07$ \citep{xie16}.

\begin{figure}
\includegraphics[width=\columnwidth]{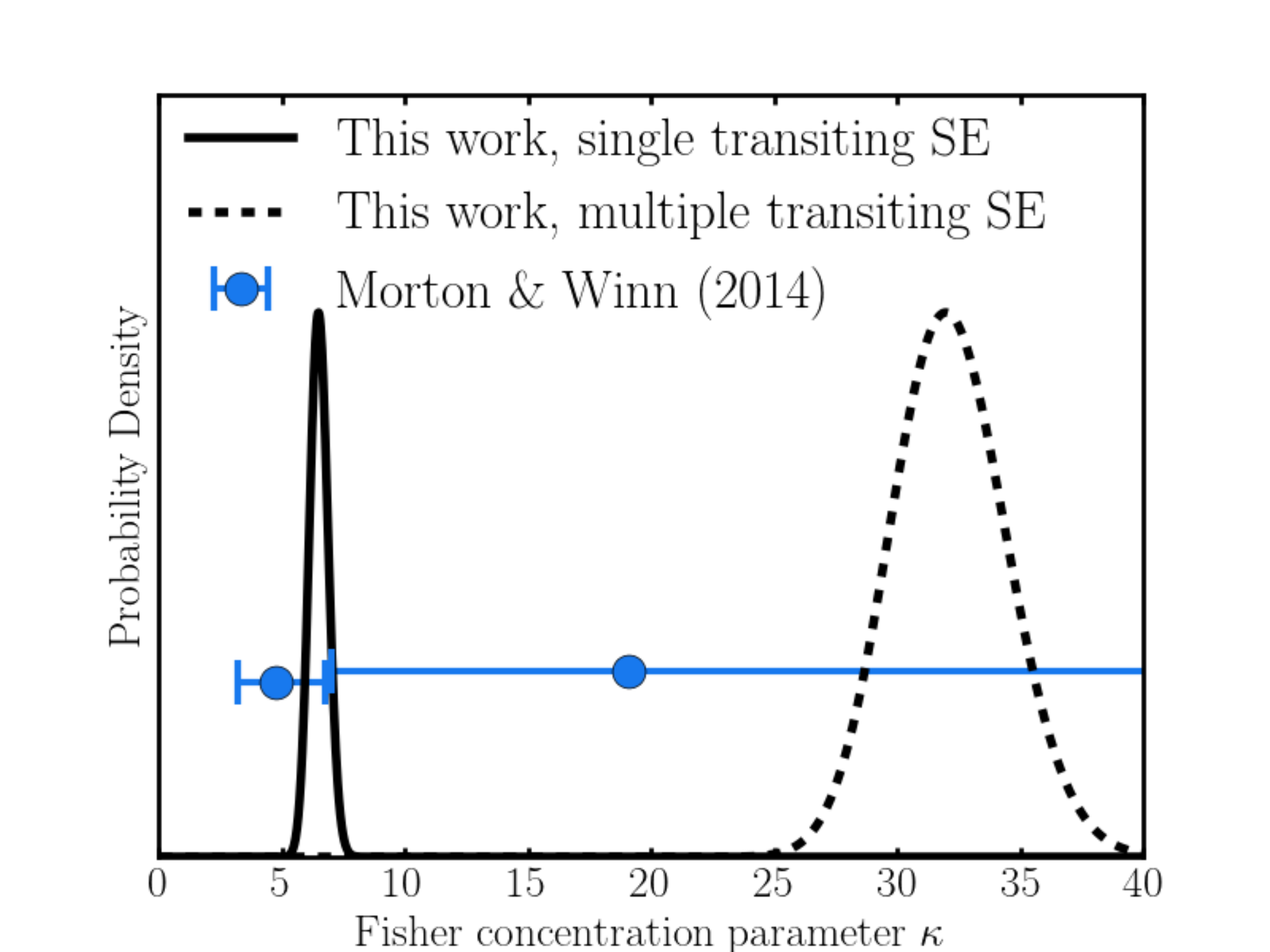}
    \caption{The posterior probability distribution of the Fisher concentration parameter $\kappa$ for the single (solid) and multiple super-Earth population (dashed) from the fiducial run. Our best fit $\kappa$ value is $6.5\pm0.4$, and $32\pm2$, respectively. To compare with, we show the values derived from \textit{Kepler} systems \citep{MW14} in blue: $\kappa=4.8^{+2.0}_{-1.6}$ for single systems, and $\kappa=19.1^{+73.4}_{-12.1}$ for multiple planetary systems. \label{fig:kappa}}
\end{figure}

We further characterize the obliquity distribution using the Fisher concentration parameter $\kappa$ following \citet{TD11} and \citet{MW14}. 
\begin{equation}
    f_{\theta}(\theta|\kappa)=\frac{\kappa}{2\sinh\,\kappa}\,\exp{(\kappa\cos\,\theta)}\,\sin\,\theta,
\end{equation}
in which, $\theta$ is the obliquity. 
We follow the method detailed in \citet{MW14} to calculate the posterior probability\footnote{We note that a uniform prior for $\kappa$ would not significantly change the result compare to the prior following \citet{MW14} Eq 3.} of $\kappa$. The result is presented in Figure \ref{fig:kappa}. We obtain best fit $\kappa$ value for our single super Earths to be $6.5\pm0.4$, which is consistent with the measurement of \citet{MW14} ($\kappa=4.8^{+2.0}_{-1.6}$) on \textit{Kepler} single systems using $vsini$ and rotation period of the host star. We also found $\kappa=32\pm{2}$ for the population with two and three super-Earths with more than one planets transits. This is tentatively similar to the $\kappa=19.1^{73.4}_{-12.1}$ derived by \citet{MW14} for the \textit{Kepler} multiple systems.  

In summary, the width of the eccentricity and stellar obliquity distributions shrink as a function of the number of Super-Earths and widen as a function of the mutual inclinations between planetary orbits. As a result, a population of single transiting super-Earths with eccentricities of $\bar{e}\sim0.3$ and obliquity angles of $\sim20^\circ$ are created.

\begin{figure*}[htbp!]
\includegraphics[width=\columnwidth]{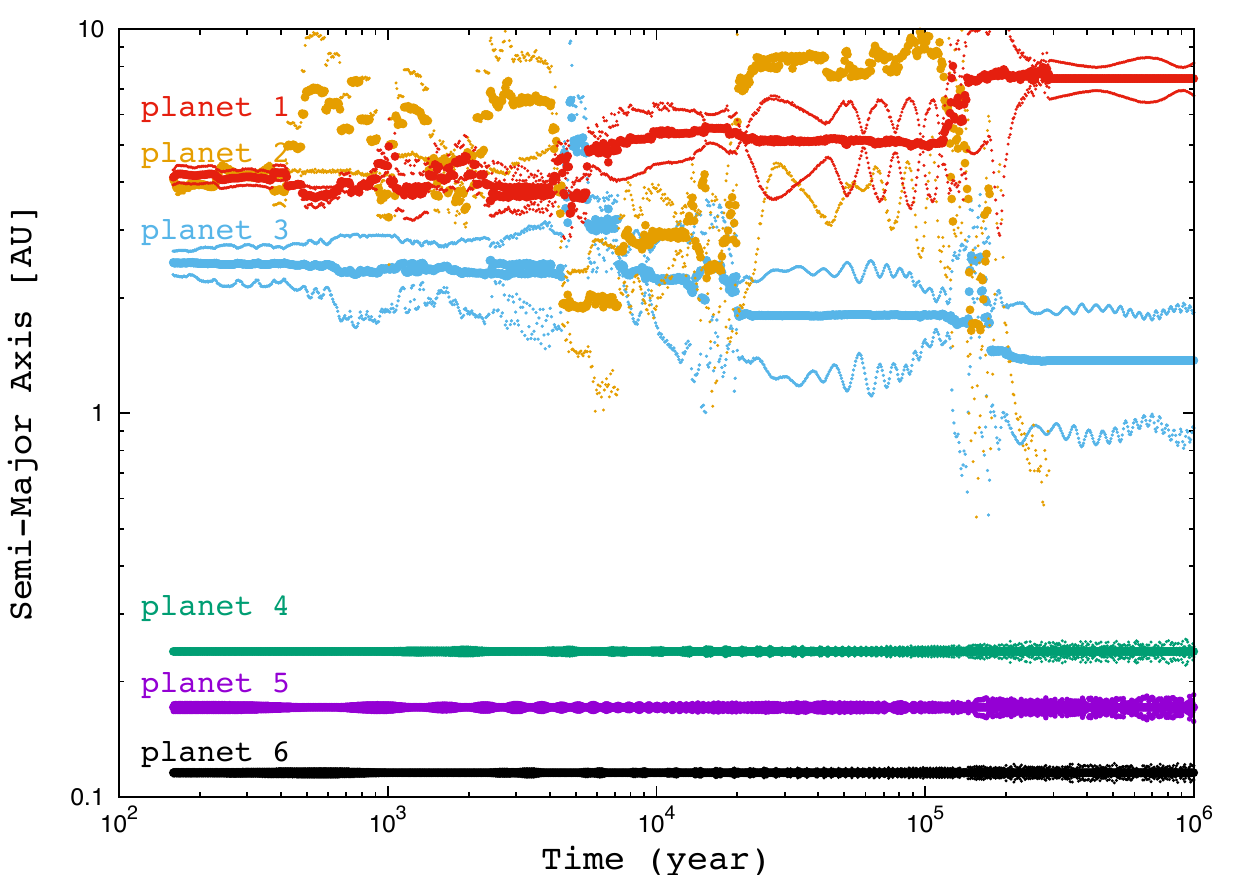}
\includegraphics[width=\columnwidth]{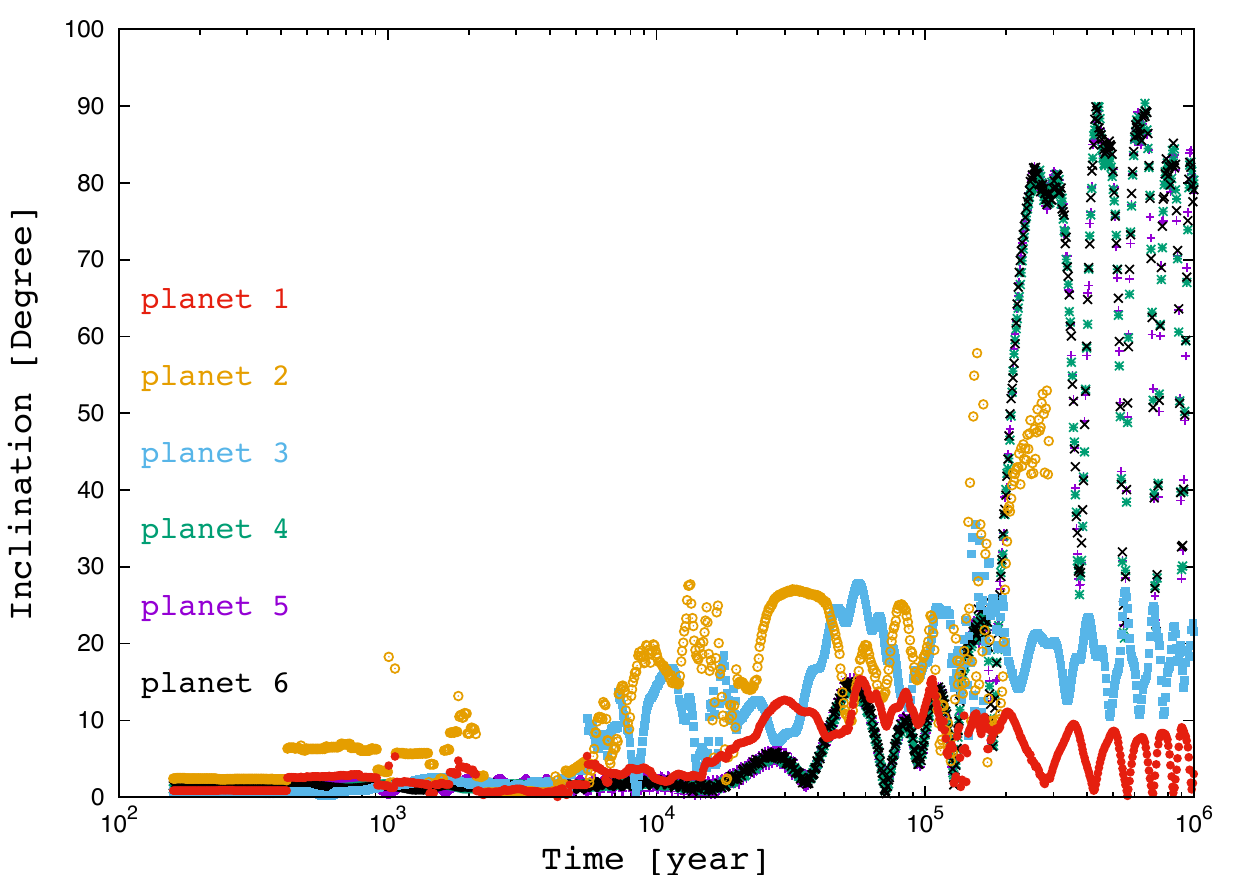}
\caption{Time evolution of a realization of our simulation which resemble the Kepler-56 system. The left panel shows how the semi major axis (thick lines), pericenter (thin lines) and apocenter (thin lines) of 6 planets vary in 1\,Myr. The right panel shows the inclination of all the planets in degrees. One of the giant planet (planet 2) is ejected at 0.2\,Myr. At the end of 1\,Myr, This system has three roughly coplanar and circular super-Earths (planet 4,5 and 6) with two eccentric giant planets (planet 1 and 3), while the inclination of the super-Earths oscillate between 30 to 90 degrees. 
\label{fig:orbit_tilt}}
\end{figure*}

Finally, we have identified a class of multiple planet systems with low mutual inclinations between the super-Earths, but large obliquity angles, similar to the Kepler-56 system (upper left region of the right panel of Figure \ref{fig:inc_earthonly}).
However, systems like these, which would violate the trend that multiple transiting systems tend to have low obliquities,  are somewhat uncommon. For reference, $\sim 1\%$ ($\sim  3.5\%$) of the systems with $N_{\rm SE}=2$ ($N_{\rm SE}=3$) have a final configuration such that $i_{\rm m}<5^\circ$ for all planet pairs, but the obliquity angles are $>20^\circ$. 
We show the time evolution of one of the Kepler-56-like systems in Figure \ref{fig:orbit_tilt}. In this example, the ejection of planet 2, a giant planet, at 0.2\,Myr has tilted the orbital plane of all the three super-Earths all together, leading to large-amplitude inclination (or, equivalently, obliquity)
oscillations between $\sim 30^\circ$ to $\sim 90^\circ$, while keeping the mutual inclinations small ($\lesssim3^\circ$). The large oscillations in obliquity are likely due to a secular resonance affect: the forcing due to the innermost Jupiter at $\gtrsim 1$ AU with timescale of $\sim10^5$ yr nearly matches a natural precession frequency of the inner planets (see also \citealt{Li2014, GF16}).

\subsubsection{Eccentricities of the outer giant planets}

\begin{figure*}[htbp!]
\centering
\includegraphics[trim=10 0 10 0, width=0.66\columnwidth]{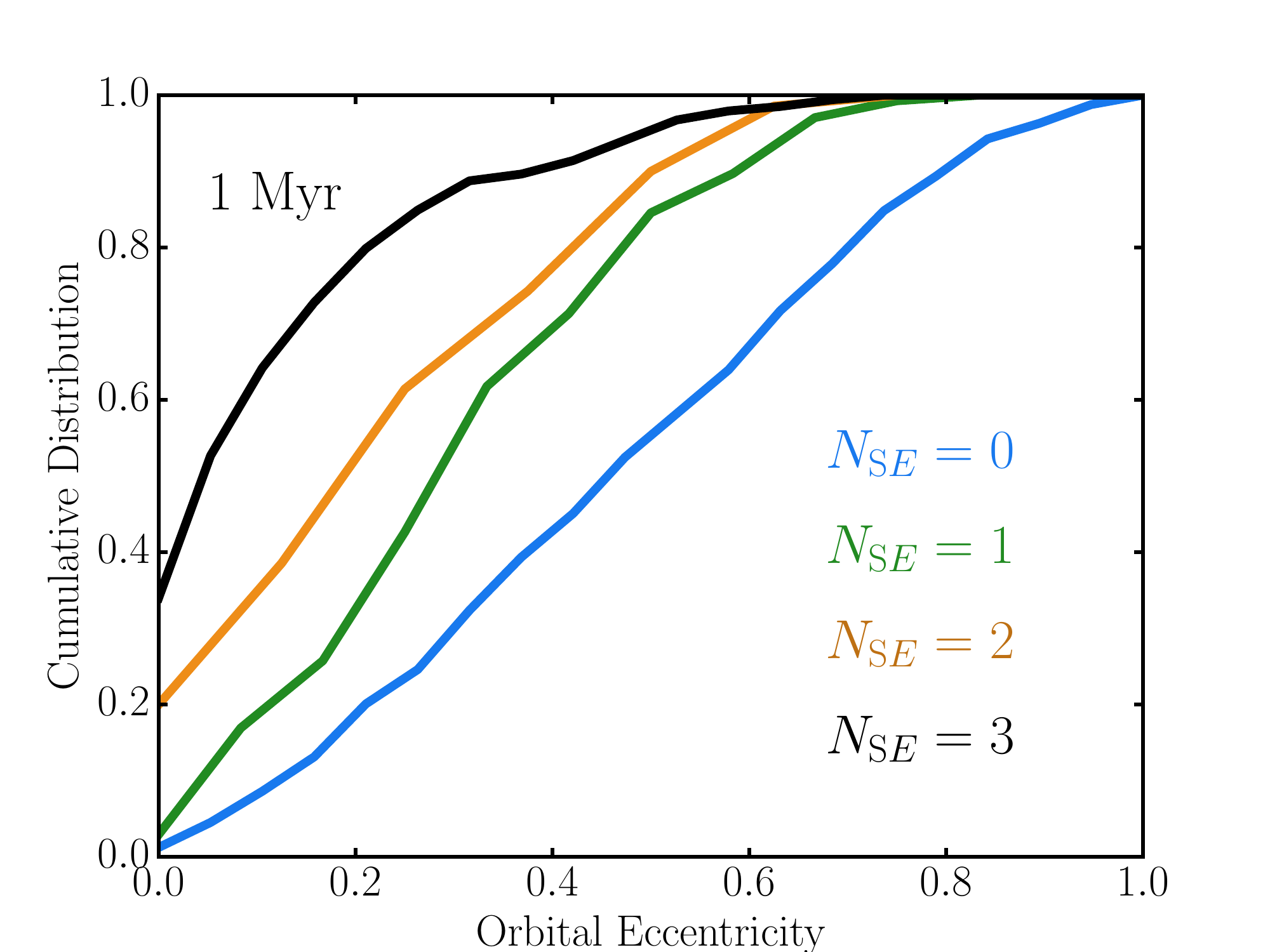}
\includegraphics[trim=10 0 10 0, width=0.66\columnwidth]{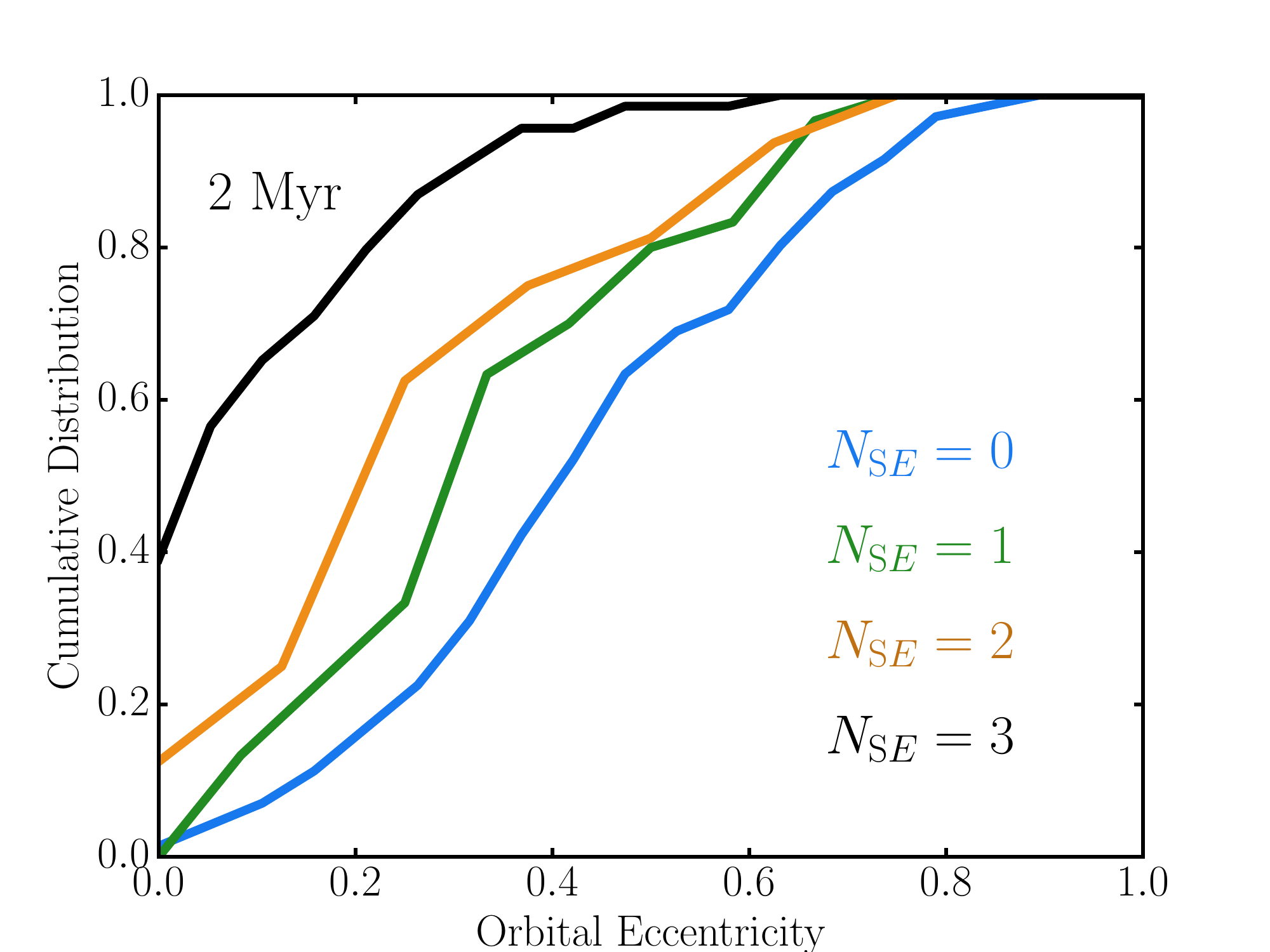}
\includegraphics[trim=10 0 10 0, width=0.66\columnwidth]{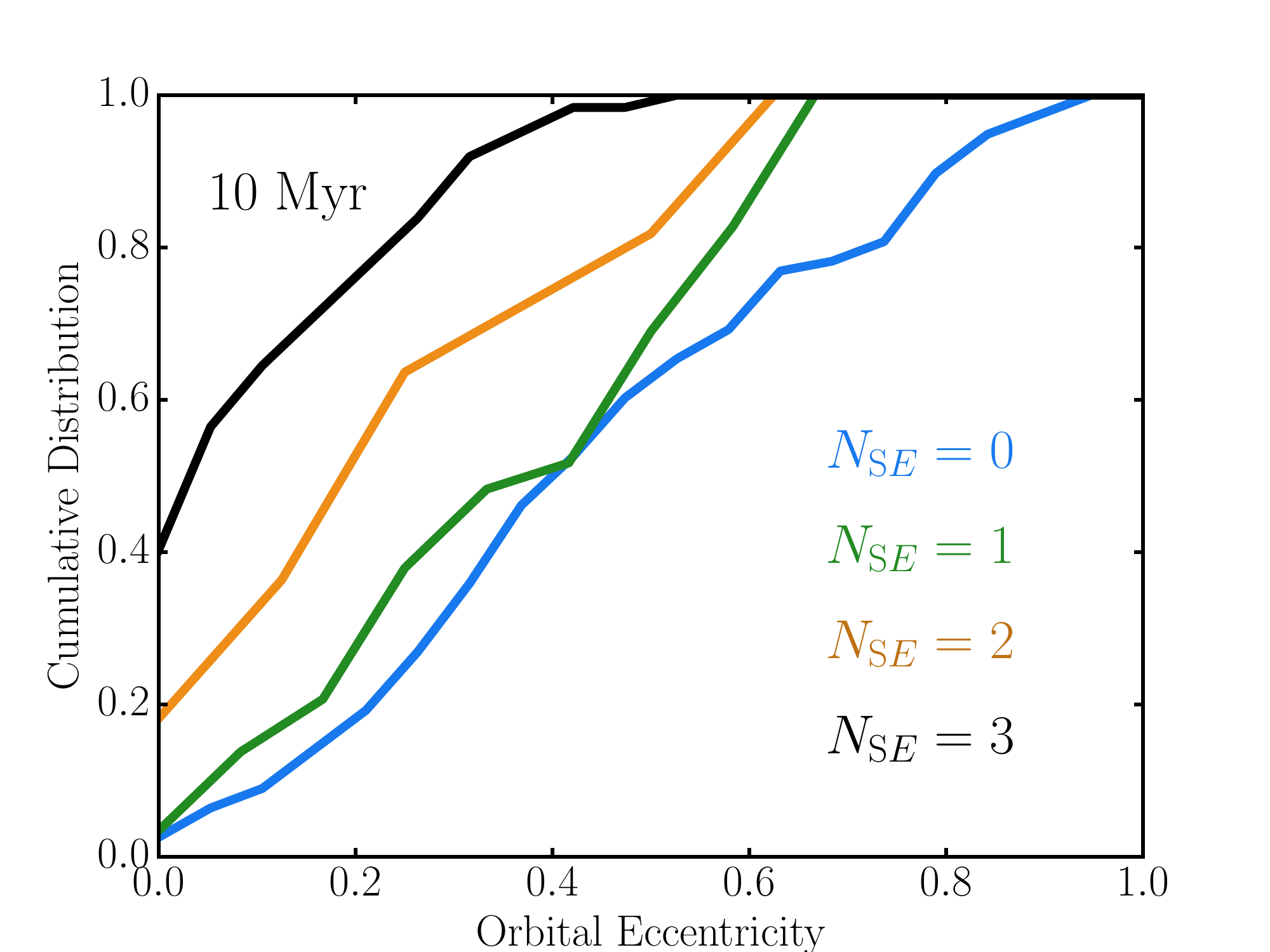}
\caption{The cumulative histogram of eccentricities for giant planets within $a<5$ AU with super-Earths in the same system. Blue, green, orange and black histograms shows the result for systems with 0, 1, 2 and 3 super-Earths remaining, respectively. From left to right, we show the result simulated for 1\,Myr, 2\,Myr and 10\,Myr time.  
\label{fig:ejup_nearth}}
\end{figure*}

We also identified the giant planets with semi major axis smaller than 5 AU, and compare their eccentricities in Figure \ref{fig:ejup_nearth}. We report that only Jupiters with eccentricities $<0.7$ can keep super-Earths in the system. The lower the eccentricity of the Jupiters is, the more likely the system is to host higher multiplicity super-Earths systems. For example, most of the three super-Earths system are likely to be found in systems where the Jupiters have eccentricities $<0.3$.

\begin{table*}[htbp!]
\centering
\begin{tabular}{c c c c c}
\hline
\hline
($\text{N}_{\text{J}}$, $\text{N}_{\text{SE}}$) & fiducial run (\%)  & GR (\%) & 2\,Myr (\%) & 10\,Myr (\%)\\
\hline
(1, 0) & 8 & 6 & 13 & 17\\
(1, 1) & 3   & 3 & 4 & 5\\
(1, 2) & 1  & 1 & 1 & 2\\
(1, 3) & 4 & 2 & 3 & 2\\
(2, 0) & 28  & 24 & 33 & 33\\
(2, 1) & 16  & 18 & 14 & 13\\
(2, 2) & 7  & 6 & 8 & 5\\
(2, 3) & 27  & 34 & 20 & 19\\
(3, 0) & 2  & 1 & 0 & 0\\
(3, 1) & 1  & 0 & 0 & 0\\
(3, 2) & 0  & 0 & 0 & 0\\
(3, 3) & 4  & 4 &4 & 3\\
\hline

\multicolumn{4}{c}{\vspace{1cm}}\\

%\end{tabular}
%\caption{Architecture of resulted planet systems.}
%\label{tab:results}
%\end{table*}

%\begin{table*}[htbp!]
%\centering
%\begin{tabular}{c c c c c c}
\hline
\hline
 & fiducial run  & GR & 2\,Myr & 10\,Myr\\
\hline
$\bar{i}$ -- $\tilde{i} $ ($^\circ$) & $30\pm2$ -- $26\pm2$   & $30\pm3$ -- $32\pm4$& $27\pm3$ -- $26\pm3$ & $28\pm5$ -- $21\pm3$\\
$\bar{e}$ -- $\tilde{e}$ & $0.39\pm0.02$ -- $0.35\pm0.04$ & $0.43\pm0.03$ -- $0.41\pm 0.04$ & $0.44\pm0.04$ -- $0.50\pm0.06$  & $0.41\pm0.04$ -- $0.40\pm0.05$\\
\hline
\end{tabular}
\caption{Upper table: Final number of outer giant planets and super-Earths for each ensemble of integrations. Lower table:  mean and median of the eccentricities and inclinations of the single super-Earth systems. }
\label{tab:results}
\end{table*}

\begin{figure}[htbp!]
\includegraphics[width=\columnwidth]{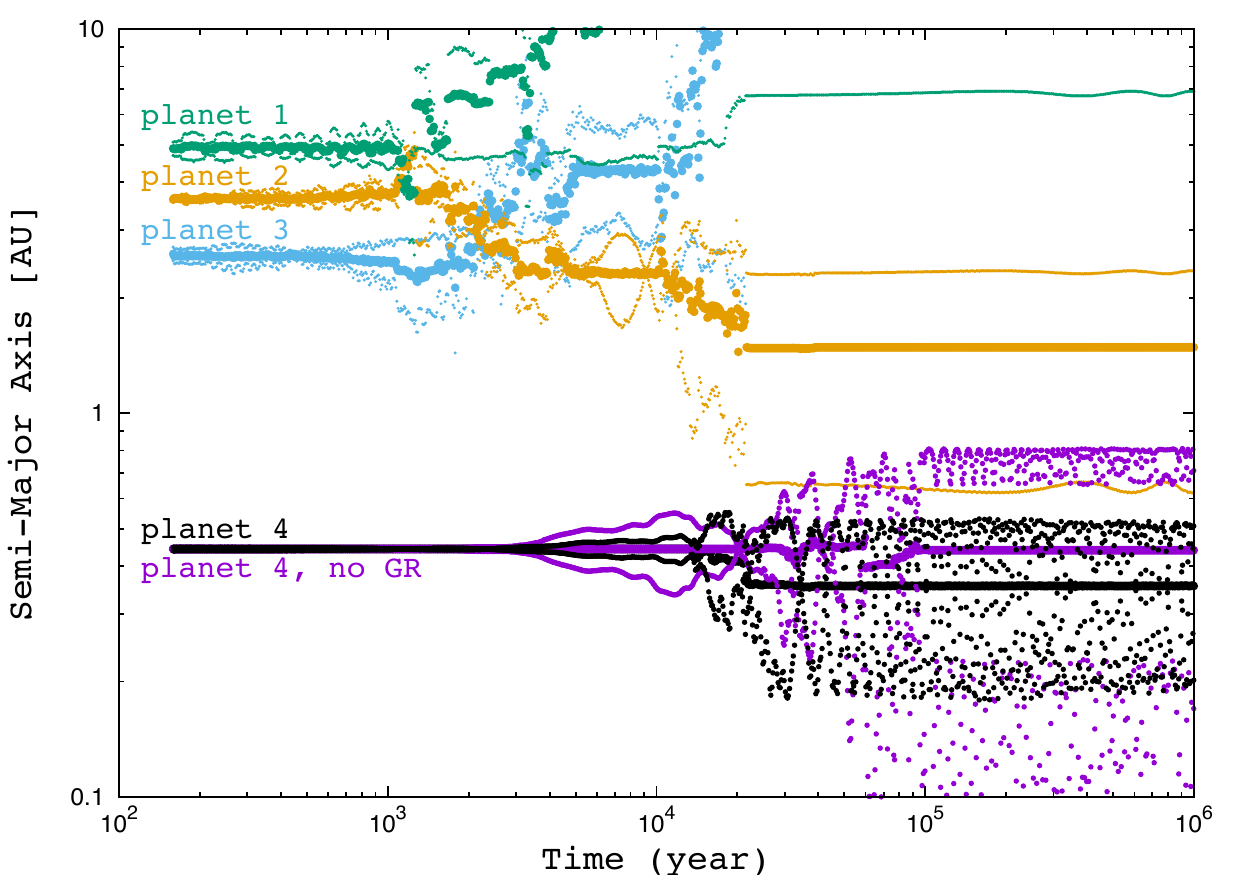}
\caption{Semi major axis (thick line), pericenter and apocenter (thin lines) of planets over time with the general relativity potential taken into account. The system start with 6 planets, with planet 1, 2 and 3 to be Jupiter like planets, and planet 4,5,6 to be close-in super-Earths. Planet 5 and 6 were ejected around $10^4$\,yrs, and are not shown in the plot. The black curve shows the survived super-Earth (planet 4) has a modest eccentricity of about 0.4. For comparison, the purple curve shows the evolution of planet 4 without the general relativity effect. The planet's eccentricity can growth to as high as 0.8 and will eventually collide with the host star without taken into account of GR.  
\label{fig:GRexample}}
\end{figure}

\subsection{Effect of General Relativity}

We include the effect that apsidal precession from general relativity has on the evolution of the super-Earths.

We simulated 160 systems with the same initial conditions as our fiducial simulations, but including  the effect from GR apsidal precession. 
The outcomes are shown in Table \ref{tab:results}. We note that
 the overall demographics  is not changed dramatically compared to our fiducial simulation.
 Looking at the results in more detail, we note that the main effect of GR is to slightly increase the number of surviving super-Earths from $60\pm3\%$ to $68\pm6.5\%$ after 1 Myr at the expense of reducing the number of systems with two giant planets and no super-Earths. This effect might be expected since the apsidal precession of the inner super-Earths can be fast enough (timescale of $\sim P_{\rm SE}\cdot(GM_\odot)/(a_{\rm SE}c^2)\sim10^5$ years, with $P_{\rm SE}$ and $a_{\rm SE}$ indicating the period and semi-major axis of the super-Earth) to quench the eccentricity excitation driven by the secular perturbations from the outer giant planets (timescale of $\sim P_{\rm SE}\cdot(M_\odot/m_J)\cdot(a_J/a_{\rm SE})^3\sim10^4-10^6$ years) that can cause the destabilization of the inner super-Earths (e.g., \citealt{MG09,WL11}).
 
 In Figure \ref{fig:GRexample}, we show one example that
illustrates how GR decreases the maximum eccentricity achieved by the single super-Earth (purples lines do not include GR and black lines do include GR) that survives the early scattering phase that removes the other two super-Earths and one giant planet. In this example the maximum eccentricity of the super-Earth is $\gtrsim0.8$ without GR, while it reaches $\lesssim0.4$ when GR is included.
We caution that our calculations ignore the orbital precession due to the tidal bulges, which can have a stronger effect than GR at limiting the maximum eccentricity that the super-Earths can reach and prevent the collisions with the host star (e.g., \citealt{WL11,P15,liu15}).
 
Consistent with the expectation that GR precession can limit the eccentricity growth due to secular perturbations and prevent that some single super-Earths collide with the host star, we observe from Table \ref{tab:results} that the number of single super-Earths with two outer giant planets ($N_{\rm J}=2$, $N_{\rm SE}=1$) increases from $\simeq13\pm1.5\%$ in the integrations without GR to $\simeq18\pm3.3\%$ in the simulations with GR.
However, even though the maximum eccentricities seem to be limited when including GR, we observe that the overall distributions in both eccentricities and inclinations are consistent (within error bars) with the other runs without GR (see Table \ref{tab:results}).

\subsection{Effect of run time}
\label{sec:time}

We analyze the long term stability of these super-Earths systems by extending 160 of our simulations for up to 2\,Myrs. As show in Table \ref{tab:results}, the fraction of systems with the super-Earths completely destroyed did not change significantly. About 6\% more systems have been destroyed in the second million year of simulation. This is mostly due to the perturbations of the systems with three super-Earths with most eccentric outer giant planets. As shown in Figure \ref{fig:ejup_nearth}, the maximum eccentricity for a giant planet to have three super-Earths companion reduced from 0.6 to 0.3 with the run time increase. In the meanwhile, the rate of single super-Earths is kept roughly constant. The final eccentricity and inclination distribution of single super-Earths can be described by $\bar{e}=0.44\pm0.04$, and $\bar{i}=27\pm3^{\circ}$. 

We integrate these simulations further for up to 10\,Myrs. 
In order to so, we use the ``wfhast" integrator, which is a fast and unbiased implementation of a symplectic Wisdom-Holman integrator for long term gravitational simulations \citep{RT15}.  
This choice is justified because the rate of close encounter between the super-Earths have reduced dramatically in the first 2\,Myrs. From these integrations we observed that only $4\%$ more of the system have been destroyed, suggesting  that the systems nearly reach an equilibrium in terms of numbers of planets. This can also be observed in the eccentricity distribution of giant planets in Figure \ref{fig:ejup_nearth}. The final eccentricity and inclination distribution of the single super-Earths can be described by $\bar{e}=0.41\pm0.04$, and $\bar{i}=28\pm5^{\circ}$. As a consequence of the reduced relative fraction of the three super-Earths system, the observed mean and median eccentricity are slightly increased ($\bar{e} = 0.21\pm0.04$, and $\tilde{e} = 0.16\pm0.04$) compared to the 1\,Myr result.

\subsection{Number of transiting systems.}

\begin{figure}[htbp!]
\includegraphics[width=\columnwidth]{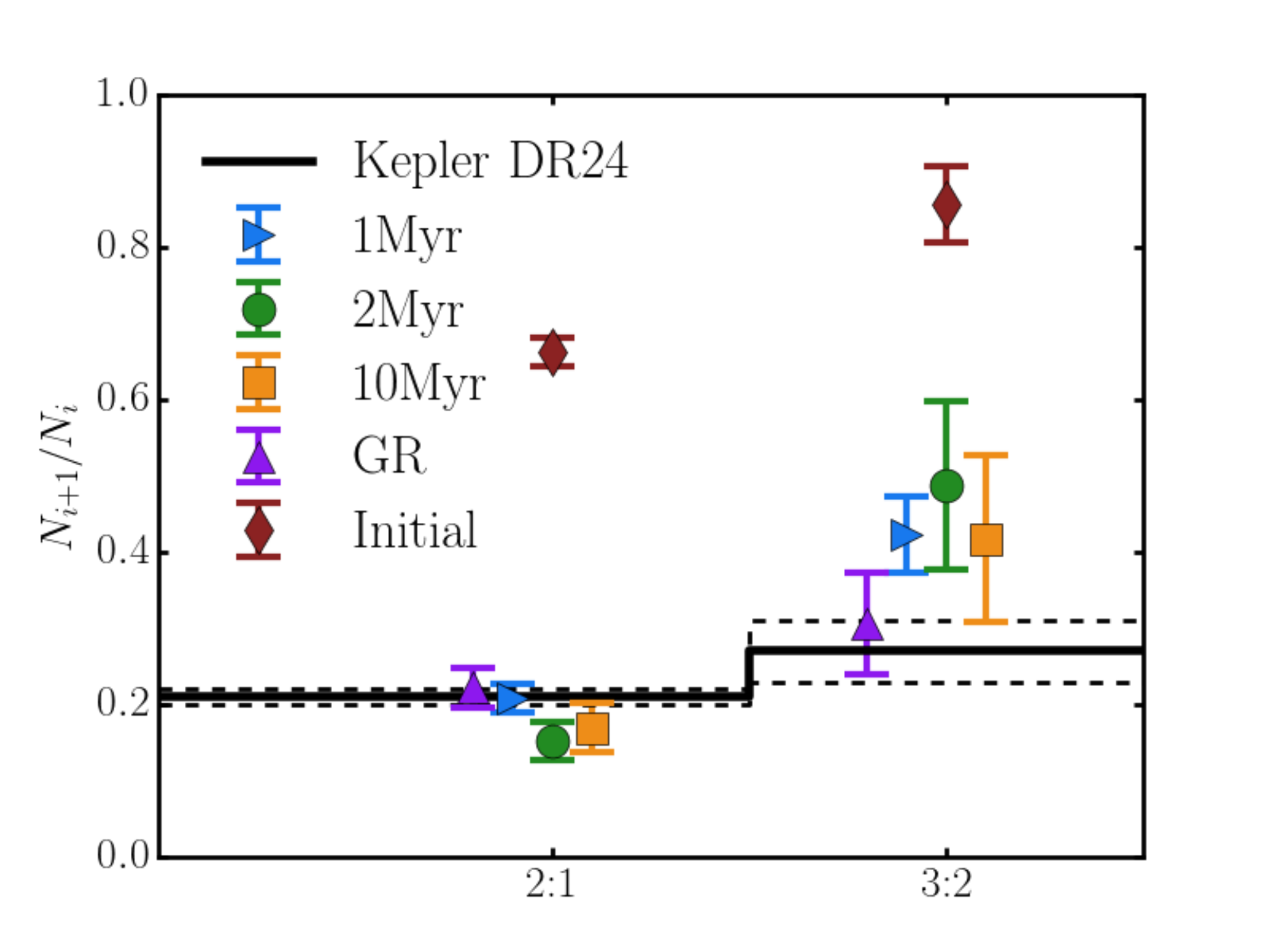}
\caption{The ratio between the number of systems with n transiting planets relative to n-1 transiting planets. We show the result at the initial of the simulation (dark red diamonds), from the 1\,Myr (blue triangles), 2\,Myr (green dots) and 10\,Myr (orange squares) evolutions. The simulation with general relativity for 1\,Myr is show in purple upper triangles. For comparison, the result from \textit{Kepler} DR24 is shown with black lines. The dash lines indicating the 1-$\sigma$ uncertainties from the \textit{Kepler} data.
\label{fig:transits}}
\end{figure}

We used the code \texttt{CORBITS} to determine the transit probability of super-Earths in the remaining systems. 
We use bootstrap to compute the average (and 1-$\sigma$ uncertainties of) numbers of systems with single transit planet, two transit planets and three transit planets.  
The ratio between observing n transiting planets relative to n-1 transiting planets is presented in Figure \ref{fig:transits}. From Data Release 24 of \textit{Kepler} candidates, we compute that the ratio between two planets system to single planet system is $0.21\pm0.01$, while the ratio between the three planets system to two planets system to be $0.27\pm0.04$. We note that although the \textit{Kepler} candidates in multiple systems are quite likely to be real planets, the single systems could be polluted by potential false positives \citep{Fressin:2013}. Thus, the intrinsic value for the ratio between two planets system and single planet systems are likely to be higher. Our initial condition is more coplanar than the \textit{Kepler} systems, with both ratios much higher. After the systems evolved for 1\,Myr, the probabilities of observing higher multiplicity systems significantly reduced, yields ratios of $0.23\pm0.02$, and $0.39\pm0.05$. As the systems evolve further, we do not see that this ratio changes significantly. The result at the end of 10\,Myr simulation yields $0.17\pm0.03$, and $0.41\pm0.11$, which are marginally consistent with the observations. 

We caution that our mechanism does not aim to explain the ``\textit{Kepler} dichotomy", namely the excess of single transiting systems. Given the relatively low occurrence rate of the giant planets ($10-20\%$, \citet{Cumming:2008, Mayor:2011}), the observed  \textit{Kepler} single transiting systems are likely to be a mixture between the perturbed and unperturbed systems.  

\section{Discussion}
\label{sec:discussion}

Our work explores the dynamical imprints that an unstable outer planetary system composed of three giant planets with final orbital configurations similar to the Radial Velocity sample can have on an inner system with three super-Earths, similar to those observed in the {\textit Kepler} sample. 

The main results of these experiments can be summarized as follows:

\begin{itemize}
\item 
The scattering events with the outer giant planets clear the inner super-Earths in nearly half of the systems.
For the remaining half of the systems, where at least one super-Earth survives, the multiplicity is greatly reduced: $\sim 30-40\%$  have only one super-Earth, while 
 $\sim 10-20\%$ have two super-Earths (see Table \ref{tab:results}).

\item The eccentricities and inclinations (or, stellar obliquity angles) of the surviving super-Earths can be excited to large values ($e\gtrsim0.3$ and $i\gtrsim20^\circ$) and their distributions widen for systems with less planets and for systems with larger mutual inclinations in the case where 2 or 3 super-Earths survived (see Figures \ref{fig:inc_earthonly} and \ref{fig:obl_nearth}).
In particular, the eccentricities and obliquity angles of the single Super-Earths roughly follow a Rayleigh distribution with $\bar{e}=0.4\pm0.02$  and $\bar{i}=30\pm2^{\circ}$, respectively. 

\item The simulations with outer giant planets decrease the relative frequency of
observing higher multiplicity systems and lower multiplicity systems in transit by a factor of 2-3. Thus, the expected number of systems with n transiting planets relative to n-1 transiting planets can better reproduce the results from {\it Kepler} (the so-called Kepler dichotomy, see Figure \ref{fig:transits}).

\item The eccentricity distribution of the outer giants shrinks as a function of number of surviving Super-Earths: the median decreases from $e\sim0.3$ for $N_{SE}=1$ (or 2) to 
to $e\lesssim0.1$ for $N_{SE}=3$. The eccentricity distribution of the systems with no super-Earth is flat in $\sim 0-0.9$.

\end{itemize}

In what follows, we discuss our results in the context of recent observations and theoretical studies.

\subsection{Implications for the {\it Kepler} sample}

As discussed in \S\ref{sec:e_i_SE}, the eccentricity and inclination distributions of the surviving super-Earths 
widen for systems with less super-Earths and for systems that 
have wider mutual inclinations, which are more 
likely to be observed as single transiting system. These predictions are in-line with recent observational evidences suggesting that \textit{Kepler} single systems are dynamically hot. In particular, the eccentricity distribution of our predicted single super-Earths has a mean of $\sim0.4$. When observing all the survived systems (single and multiple systems together) randomly, we estimated that the single transiting planets have mean eccentricities of $\sim0.2$. This is somewhat lower than what it has been found by \citet{xie16} for \textit{Kepler} single systems ($\bar{e}\simeq0.3$). We caution that the observed eccentricity distribution of transiting single super-Earths from our simulation can be quite sensitive to the relative fraction between the true single and triple super-Earth systems, which is arbitrarily affected by starting every systems with three planets and might change as the systems evolve for longer time scales. 

Similarly, by fitting a Fischer distribution with parameter $\kappa$ we find that the best fit to the obliquity distribution of the single systems is $\kappa\sim6.5$ (Figure \ref{fig:kappa}, solid line), compared to $4.8^{+2}_{-1.6}$ found in \citet{MW14}. A similar calculation for the population of multi-planet systems that shows multiple transits, yield $\kappa\sim32$ (Figure \ref{fig:kappa}, dashed line), compared to $19.1^{+73.4}_{-12.1}$ found in \citet{MW14}. 

Our experiment predicts that long term Radial Velocity follow up of \textit{Kepler} single systems will likely yield a population of giant planets beyond $\sim1$ AU, especially around those systems with transit durations that are unlikely to be explained by circular orbits. It would be difficult, however, for the \textit{Kepler} sample to measure the rate of such a mechanism happen, due to the relative faint nature of the planet hosting stars. Known Radial Velocity systems such as HD\,125612 show this type of architecture are not rare. We note that majority of the multiple super-Earth systems discovered by Radial Velocity surveys are likely to have similar mutual inclination distribution compared to the Kepler multiple systems \citep{Figueira:2012}. However, the obliquity distribution of the single Radial Velocity super-Earths are vastly unknown. Future discoveries from \textit{TESS} and follow up studies will reveal the prevalence of this mechanism in shaping the dynamical properties of the single transiting planets.

Our model can reproduce the observed ratios of multiplicities in the \textit{Kepler} sample (Figure \ref{fig:transits}) and, therefore, it provides some contribution to the traditional ``\textit{Kepler} dichotomy" problem. 
However, our model, which relies on the presence of giant planets, can hardly account for the ``dichotomy" simply because there are not enough giant planets (relative to inner super-Earths) to make up for a sizable contribution. 
 Moreover, the ``\textit{Kepler} dichotomy" problem persists in M stars \citep{Ballard2016}, where giant planets are more infrequently found than in F and G stars \citep{Johnson2010}, although microlensing surveys indicates that this problem maybe less severe \citep{Clanton:2014}.

\subsection{Comparison to other works}

There has been a wealth of recent theoretical work devoted to study the origin of the orbital architecture of the \textit{Kepler} planets.
In what follows, we discuss these works in relation of ours. We separate these works in two broad categories: with and without outers perturbers.

\subsubsection{No outer massive perturbers}

One possibility is that the excitation of eccentricities and inclinations occurs during the assembly process itself and its subsequent long-term evolution.

 \citet{HM13} and \citet{T15} have studied the predictions from a model in which planets form after the gas disk has dissipated by mergers of embryos (giant impact phase). These works predict an eccentricity distribution following an exponentially decaying function $p(e)=e/\bar{e}  \exp(-e/\bar{e})$, while the simulations of \citet{HM13} find $\bar{e}=0.057$. Such distribution might explain the eccentricities of the multiple transiting planets in the \citet{xie16} sample, but can hardly explain the large eccentricities ($\bar{e}\sim0.3$) of the single transiting planets.
 A similar approach to these studies of planet formation has also been carried by \citet{Moriarty2015}, in which by varying the mass distribution of the embryos, the authors can match some of the bulk the orbital configurations of the Kepler systems. However, their calculations seem not excite the eccentricities to the levels required to explain the hot population of single transiting systems in \citet{xie16}.

Similarly, \citet{Pu2015,VG15} argued that there is evidence that the high multiplicity ($N_p>3$) \textit{Kepler} systems are currently at the edge of stability suggesting that these systems might have become unstable in the past losing planets. Even though the dynamical instability can reduce the multiplicity of the multiple planet systems, the unstable planets will most likely lead to planet mergers, not ejections, which would not effectively excite the eccentricities and inclinations (e.g., \citealt{Johansen2012,PTR14,MNI15}).
Also, the self-excitation of mutual inclinations in the compact systems is generally unable to bring planetary orbits out of transit and account for the excess of single transiting planets \citep{BA16,hansen2016}.

In summary, these previous studies suggest that either the assembly process of the multiple planet systems or  their long-term evolution can only modestly excite the eccentricities and inclinations, and are unlikely to account for the population of eccentric single transiting planets.

\subsubsection{With outer massive perturbers}

Similar to our work, recent studies have invoked outer massive perturbers to shape the orbital configurations of the inner super-Earths. 

\citet{Lai2016} studied whether an external inclined (relative to the super-Earths) planet or star could excite the mutual inclinations of the multiple planet system, helping to account for the large number of single transiting planets.
Although some of the outer perturbers in our experiments can excite the mutual inclinations of the inner super-Earths\footnote{The innermost outer giant planet lies at $\sim1-2$ AU (Figure \ref{fig:mass-final}), while the super-Earths are typically at $\sim0.001-1$ (Figure \ref{fig:init-pratio}) meaning that the rigidity parameter 
$\epsilon\sim m_J/m_{SE}(a_{SE}/a_J)^3\sim0.01-1$.}, their inclinations relative to the super-Earths that survive with 3 planets are typically relatively low ($\lesssim 5^\circ$) to excite large mutual inclinations.
Moreover, we note that the inclinations of the outer Jupiters anti-correlate with the number of surviving super-Earths (see Figure \ref{fig:results}), suggesting that scattering events that excites the largest inclinations of the outer giant planets efficiently removes the super-Earths, possibly due to large eccentricity excitation of the Jupiters. 

Similar to \citet{Lai2016}, \citet{hansen2016} studied the effect of secular perturbations from outer giant planets in either inclined and/or eccentric orbits. He notes that both the excitation of large inclinations and the reduction in the numbers of super-Earths can be efficient enough to account for the observed multiplicities for dynamically hot outer giant planets. 
In particular, his calculations with two giant planets show that these perturbations can efficiently remove most super-Earths, which might be related to the high efficiency of super-Earths removal in our experiments because the evolution of these planets after the ejection of one giant planet is mostly driven by the secular interactions with the giant planets. 

The strong scattering events of the giant planet can produce highly eccentric planets ($e\gtrsim0.8$) that can strongly interact the inner \textit{Kepler}-like system. This idea has been explored by \citet{MDJ15} by placing the planets in eccentric orbits, crossing those of the super-Earths, and noted that this could either reduce the number of super-Earths in the system or change the semi-major axes of the Jupiter dramatically forming a warm Jupiter or ejecting it. We expect that our experiments mostly lead to the disruption of the inner planetary systems because the binding energies of the Jupiters exceed those of the super-Earths by a factor of $\sim m_J/(3m_{\rm SE})\times(a_J/a_{\rm SE})\sim 1-10$. 
We caution, however, that we have not disentangled whether the reduction of planets in our experiments is due to close encounters with the giant planets like in the experiments by \citet{MDJ15} or by secular interactions like those in \citet{hansen2016}.

Similar to our work, \citet{GF16} recently studied the effect of outer giant planet scattering on inner multiple planet systems. Unlike our work, the authors focused on the  Kepler-56 system and studied whether scattering can explain its large stellar obliquity
($\gtrsim45^\circ$, \citealt{Huber:2013}) and low mutual inclination (see Figure \ref{fig:inc_earthonly}). The authors show that their experiments with three outer giant planets produce large enough obliquity angles, while retaining low mutual inclinations between the Kepler-56 b and c. 
Our experiments also produce a population of Kepler-56-like systems with large obliquity angles and low mutual inclinations (see cluster of two- and three-planet systems around Kepler-56 in  Figure \ref{fig:inc_earthonly}). 
We note that our experiments consider inner super-Earths with masses of $\sim10M_{\oplus}$, which are much smaller than those in the Kepler-56 system (b and c have masses of $\simeq 21M_{\oplus}$ and $\simeq 170M_{\oplus}$). Therefore, we expect that our experiments lose super-Earths or excite mutual inclinations much more frequently than those experiments in \citet{GF16}.

Most recently, \citet{MDJ16} have carried out a similar suite of experiments than ours and studied, among other things,  how outer giant planet scattering shapes the architectures of \textit{Kepler}-like inner planets.  Their calculations show that $\sim20-40\%$ of the inner systems reduce their multiplicity, which is different from our results in which $\sim70-80\%$ of the systems lose at least one planet (see Table 2). In turn, the surviving planets in their calculations have significantly lower eccentricities and inclinations than those in our calculations. We mainly attribute these differences to the choice of the inner systems, which in 
\citet{MDJ16} these are typically more resilient (typically more compact and closer-in) to the perturbations from the outer giant planets than our systems.
Similarly, part of the differences are due to our choice of the outer giant planets, which have a narrower mass ratio distribution compared to \citet{MDJ16}, possibly leading to more violent and disruptive scattering events that are reflected in our wider final eccentricity distribution of the giant planets (see \citet{Carrera:2016} for a more quantitative study of this effect). We recall that, unlike \citet{MDJ16}, we have chosen the range of masses of the giant planets such our experiments reproduce observed eccentricity distribution.

In summary,  consistent with our findings, recent works show that the effect from having outer giants can dramatically change the orbital architecture of the inner super-Earths. The final outcomes in different studies, however, depends critically on the compactness of the inner systems.

\section{Conclusion}
\label{sec:conclusion}

We run N-body experiments to study the effect that outer ($\gtrsim1$ AU) giant planet companions can have on the orbital configurations of close-in super-Earth multiple systems.  
We show that, the planet-planet scattering events that shapes the giant planets to have final orbital states that resemble those of the systems discovered by Radial Velocity surveys, can excite the eccentricity and inclination of the super-Earths. As a result, about half of the inner multiple systems are completely destroyed, with a fraction of the remaining systems have their multiplicity reduced, producing a population of dynamically hot single super-Earth systems.

We predict these single super-Earths to have mean eccentricities of $\sim0.4$ and mean inclinations $\sim 30^\circ$. As the multiplicity increases, the systems have lower eccentricity and mutual inclinations. The obliquity distribution from this mechanism agrees with the tentative observation that single transiting systems have a wider distribution of stellar obliquity angles compared to the multiple transiting systems \citep{MW14}.

\section*{Acknowledgements}
The Dunlap Institute is funded through an endowment established by the David Dunlap family and the University of Toronto. 
This project is the result of the 2016 Summer Undergraduate Research Program (SURP) in astronomy \& astrophysics at the University of Toronto. 
Simulations in this paper made use of the REBOUND code which can be downloaded freely at \href{http://github.com/hannorein/rebound}{github}. 
We are grateful for the discussion with Daniel Tamayo, Yanqin Wu and Hanno Rein. 
We thank the anonymous referee for his/her thoughtful comments. This work also benefits from generous comments from D. Fabrycky, D. Lai, J. W., Xie, and P. Figueira. 
C.P. acknowledges support from
the Gruber Foundation Fellowship.
References to exoplanetary systems were obtained through the use of
the paper repositories,
\href{http://adsabs.harvard.edu/abstract_service.html}{ADS} and
\href{http://arxiv.org/archive/astro-ph}{arXiv}, but also through
frequent visits to the \href{http://exoplanet.eu}{exoplanet.eu}
\citep{Schneider:2011} and
\href{http://exoplanets.org}{exoplanets.org} \citep{Wright:2011}
websites.  This paper includes data collected by the Kepler
mission. Funding for the Kepler mission is provided by the NASA
Science Mission directorate.  This research has made use of the NASA Exoplanet Archive,
which is operated by the California Institute of Technology, under
contract with the National Aeronautics and Space Administration under
the Exoplanet Exploration Program.

%%%%%%%%%%%%%%%%%%%%%%%%%%%%%%%%%%%%%%%%%%%%%%%%%%%%%%%%%%%%
% BIBLIOGRAPHY %
%%%%%%%%%%%%%%%%%%%%%%%%%%%%%%%%%%%%%%%%%%%%%%%%%%%%%%%%%%%%

\end{document}